\documentclass[11pt]{article}
\usepackage{jheppub}
\usepackage{pix}
\usepackage{epsfig}
\usepackage{amsfonts}
\usepackage{amsmath}
\usepackage{amssymb}
\usepackage{bbm,bm}
\usepackage{caption,subcaption}
\usepackage[normalem]{ulem}

\def\neff{N_\mathrm{eff}}
\def\kmin{k_{\mathrm{min}}}
\def\kmax{k_{\mathrm{max}}}
\def\tmin{T_{\mathrm{min}}}
\def\tmax{T_{\mathrm{max}}}
\def\dd{\mathrm{d}}
\def\dv[#1][#2]{\frac{\dd #1}{\dd #2}}
\def\ddv[#1]{\frac{\dd}{\dd #1}}
\def\pdv[#1][#2]{\frac{\partial #1}{\partial #2}}
\def\eax{\tilde{e}_a}
\def\neq{n_{\mathrm{eq}}}
\def\eeq{e_{\mathrm{eq}}}
\def\fPQ{f_\rmii{PQ}}
\def\fmin{\fPQ^\mathrm{min}}

\newcommand{\rmii}[1]{{\mbox{\tiny\rm{#1}}}}
\def\nB{n_{\rmii{B}}}
\def\nF{n_\rmii{F}}
\newcommand{\mD}{m_\rmii{D}}
\newcommand{\qm}{q_-}
\newcommand{\qp}{q_+}
\newcommand{\qmp}{q_\mp}
\newcommand{\qpm}{q_\pm}

\newcommand{\jg}[1]{#1}
\newcommand{\kb}[1]{#1}

\title{Thermal axion production at hard and soft momenta}
\author[a]{Killian Bouzoud,}
\author[a]{Jacopo Ghiglieri}

\affiliation[a]{
SUBATECH, Nantes Universit\'e, IMT Atlantique, IN2P3/CNRS,\\
4 rue Alfred Kastler,
La Chantrerie BP 20722, 44307 Nantes, France}
\emailAdd{killian.bouzoud@subatech.in2p3.fr}
\emailAdd{jacopo.ghiglieri@subatech.in2p3.fr}

\abstract{
Hot axions, thermally produced in the Early Universe, would contribute to dark radiation and are thus
subject to present and future constraints from $\neff$. 
In this paper we quantify the contribution to $\neff$ and its uncertainty in models with axion-gluon couplings
from thermal dynamics above the QCD transition. 
In more detail, we determine the leading-order thermal axion production rate 
for axion momenta of the order of the temperature adopting three different schemes for the 
incorporation of the collective dynamics of soft gluons. We show how these three schemes extrapolate
differently into the regime of softer axion production, thus giving us a \kb{first} quantitative handle on the theory
uncertainty of the rate. 
Upon solving the Boltzmann equation, we find that this theory uncertainty translates to
an uncertainty of \kb{order} 0.002 for \kb{the contribution to} $\neff$ \kb{prior to the QCD 
crossover}. The  uncertainty from
common momentum-averaged approximations to the Boltzmann equation is smaller.
We also discuss how
QCD transition dynamics would need to be integrated into our results \jg{and we show
how existing rate determinations in the literature based on gauge-dependent resummations are problematic.}
}

\begin{document}
\maketitle

\section{Introduction}
The axion was introduced  in \cite{pecceiquinn,weinberg,wilczek}
as a simple, elegant solution to the so-called strong CP problem of Quantum Chromodynamics (QCD). 
It was later realized in \cite{Preskill:1982cy,Abbott:1982af,Dine:1982ah} that such a new field could account for the observed amount of dark matter --- see \cite{Marsh:2015xka,DiLuzio:2020wdo} for reviews. 
Here we rather focus on another cosmological consequence: the couplings between 
the axion and the Standard Model (SM) particles cause \emph{thermal production} of \emph{hot axions}
during the radiation epoch of the early universe.

In a nutshell, for temperatures $T$ sufficiently lower than the axion scale $\fPQ$,\footnote{%
We use this somewhat nonstandard notation, where PQ stands for Peccei--Quinn, rather than the more
common $f_a$, because later on we shall reserve that symbol for the axion phase-space distribution.
}
axion-SM interactions are encoded by non-renormalizeable operators suppressed 
by powers of $T/\fPQ$. Hence, at sufficiently high temperatures ultrarelativistic axions will be in equilibrium
with the other bath constituents, to later \emph{freeze out}. This frozen-out abundance constitutes
\emph{dark radiation} and is thus constrained by the number $\neff$ of light degrees of freedom at the epochs
of Big Bang Nucleosynthesis (BBN) and of  Cosmic Microwave Background (CMB) decoupling. Therefore, determinations of 
the frozen-out abundance can be used, together with observational constraints, to probe parameters of axion modes,
chiefly $\fPQ$ itself. We mention that Planck data \cite{Planck:2018vyg} yield $\neff=2.99\pm 0.17$ and combined BBN data give $\neff=2.898 \pm 0.141$ \cite{Yeh:2022heq},
from which any particle degree of freedom having frozen out around or after the QCD crossover transition 
at $T_c\approx 150$ MeV can already be excluded.

These observational limits are expected to improve by an order of magnitude with future 
detectors~\cite{CMB-S4:2016ple,Abazajian:2019eic,CMB-S4:2022ght}, making higher freeze-out temperatures accessible
and approaching the uncertainty of existing high-precision theory determinations of $\neff$ in the 
SM~\cite{Bennett:2019ewm,Bennett:2020zkv,Cielo:2023bqp,Drewes:2024wbw}. 
This then motivates a renewed look at calculations of the interaction rate of axions in the early
universe, which is the key ingredient for the determination of the frozen-out abundance. It is defined as 
\begin{equation}
    \label{defintrate}
    \partial_t f_a(k)-H k\, \partial_k f_a(k)=\Gamma(k)\big[\nB(k)-f_a(k)\big] \,,
\end{equation}
where $f_a(k)\equiv(2\pi)^3 \mathrm{d} N_a/(\mathrm{d}^3\mathbf{k}\mathrm{d}^3\mathbf{x})$ is the axion
phase-space density and $\nB(k)\equiv(\exp(k/T)-1)^{-1}$ its Bose--Einstein form in equilibrium. 
As the axion mass is negligible for hot axions, $E_k\approx k$, where $k\equiv \vert \mathbf{k}\vert$ is the modulus of their three-momentum. $H$ is the Hubble rate. 
As the axion production rate $\Gamma(k)$ depends on the momentum, the notion of a freeze-out temperature is meaningful 
only when axion decoupling is fast enough that all modes freeze out in a narrow temperature range, well
approximated by a single one. This was already discussed in~\cite{Notari:2022zxo,Bianchini:2023ubu}: we  will
address this issue quantitatively and discuss similarities with the recent analysis in~\cite{DEramo:2023nzt}. 

Our main goal, however, is the precise determination of $\Gamma(k)$ itself.
We will concentrate 
on the KSVZ model~\cite{Kim:1979if,Shifman:1979if} where the axion can only interact with gluons.
Calculations of $\Gamma(k)$ have a long history, starting from \cite{Braaten:1991dd,Masso:2002np}, leading to the existing 
perturbative determinations 
and phenomenological consequences in \cite{Graf:2010tv,Salvio:2013iaa,DEramo:2021psx,DEramo:2021lgb,DEramo:2022nvb}. As we shall
illustrate, gluon-mediated processes are very sensitive to the collective effects of a QCD 
plasma, which perturbatively first appear at momenta (wavelengths) smaller (larger) than $g_3T$ ($1/g_3T$),
where $g_3$ is the QCD gauge coupling. These collective effects cure would-be infrared (IR) divergences
of the rate; their implementation requires resummed propagators and vertices, such as 
those arising from the Hard Thermal Loop (HTL) effective theory~\cite{Braaten:1989mz,Braaten:1991gm}.
As the separation between the \emph{soft scale} $g_3T$ and the 
\emph{hard scale} defined by the temperature becomes blurred once $T$ approaches $T_c$, 
the reliability of these perturbative calculations can be called into question, 
see~\cite{Notari:2022zxo,Bianchini:2023ubu}.

Here we shall discuss in detail the physics responsible for the soft-gluon contribution to $\Gamma(k)$
and explain how it is dealt with in existing calculations. We shall also explain how 
these calculations become \emph{extrapolations} once \emph{soft axion momenta}, $k\lesssim g_3T$, are considered, giving rise to
issues of negativity and gauge dependence. We shall then introduce two prescriptions which are novel 
in the context of axion production, inspired respectively by previous work on right-handed-neutrino
and gravitational-wave production \cite{Besak:2012qm,Ghiglieri:2016xye,Ghiglieri:2020mhm} and by 
QCD kinetic theory implementations \cite{York:2014wja}. We shall show how both 
are gauge invariant and how the latter in particular stays positive definite by construction
under extrapolation to $k\lesssim g_3T$. \jg{We shall also discuss 
the pathologies that arise from the gauge-dependent resummation method 
of \cite{Rychkov:2007uq,Salvio:2013iaa}.}

\kb{Following methods in use in the hot QCD literature --- see App.~\ref{app_QCD} ---} we will be able to give a \kb{first} estimate 
of the uncertainty of the rate \kb{by comparing our determinations and previous ones}. 
\kb{We will show how this uncertainty} translates quantitatively on the freeze-out mechanism
and the resulting abundance. As we mention, we shall also quantify the uncertainty related to 
adopting momentum-integrated approximations as opposed to our momentum-dependent implementation
of Eq.~\eqref{defintrate}.

The paper is organized as follows: in Sec.~\ref{sec:irdivergence}
we present a pedagogical introduction to the emergence of collective effects
in calculations of axion production, showing how they are incorporated, with their limitations. We then 
introduce our prescriptions. In Sec.~\ref{sec_prodrateres} we present numerical results for 
the rate resulting from the new prescriptions and compare them with existing determinations, paving the 
way for Sec.~\ref{sec_momdep}. There, we insert these rates in the Boltzmann equation~\eqref{defintrate} and
present our solutions with full accounting of momentum dependence. We then proceed to quantify 
the uncertainties stemming from the rate and those arising from the momentum-averaged approximations.
Finally, in Sec.~\ref{sec_concl} we draw our conclusions. Details on the technical implementation of the rates 
and on the comparison with the literature are to be found in the appendices\kb{, together with a brief summary of 
hot QCD methods and results for the assessment of the theory uncertainty of perturbative calculations}.

\section{Collective effects, scale separation and Hard Thermal Loop resummation}\label{sec:irdivergence}
Let us start by briefly discussing the mechanisms that may keep hot axions in thermal 
equilibrium at some early point in the thermal history of the universe. At energies
well below the axion scale $\fPQ$, the most
generic axion-SM effective Lagrangian reads --- see e.g.~\cite{Salvio:2013iaa}
\begin{align}
    \mathcal{L}=&\,\mathcal{L}_{\mathrm{SM}}+\frac{1}{2}(\partial_\mu a)(\partial^\mu a)
    -\frac{a}{\fPQ}\left[c_3\frac{\alpha_3}{8\pi}G^b_{\mu\nu}\tilde{G}^{\mu\nu}_b+c_2\frac{\alpha_2}{8\pi}W^j_{\mu\nu}\tilde{W}^{\mu\nu}_j+c_1\frac{\alpha_1}{8\pi}B_{\mu\nu}\tilde{B}^{\mu\nu}\right]\nonumber
    \\
    &+ic_t y_t\frac{a}{\fPQ}\bar{Q}_L\sigma_2H^{*}u_R+\text{h.c.}\,,\label{eq:laxion}
\end{align}
where $a$ is the axion field, $\alpha_i=g_i^2/4\pi$ with the $g_i$ being the coupling constants of the different gauge groups and $\tilde{T}^{\mu\nu}=1/2\varepsilon^{\mu\nu\rho\sigma}T_{\rho\sigma}$ is the dual of the tensor $T$, with $G$, $W$ and $B$
the SU(3), SU(2) and U(1) field-strength tensors.  h.c. denotes the Hermitean conjugate, $\sigma_2$
is a Pauli matrix and $y_t$ is the top Yukawa coupling,. All other SM Yukawa couplings can be considered negligible.

\begin{figure}[t]
    \centering
    \begin{subfigure}{0.24\textwidth}
        \centering
        \includegraphics[width=\textwidth]{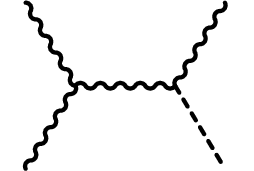}
    \end{subfigure}
    \hfill
    \begin{subfigure}{0.24\textwidth}
        \centering
        \includegraphics[width=\textwidth]{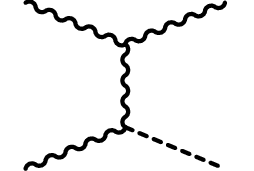}
    \end{subfigure}
    \hfill
    \begin{subfigure}{0.24\textwidth}
        \centering
        \includegraphics[width=\textwidth]{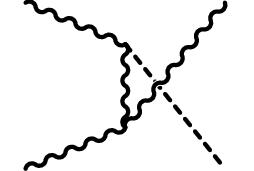}
    \end{subfigure}
    \hfill
    \begin{subfigure}{0.24\textwidth}
        \centering
        \includegraphics[width=\textwidth]{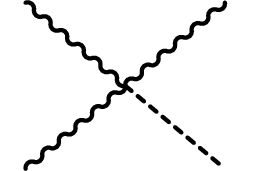}
    \end{subfigure}
    
    \bigskip
    \begin{subfigure}{0.24\textwidth}
        \centering
        \includegraphics[width=\textwidth]{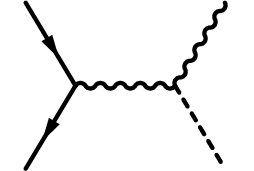}
    \end{subfigure}
    \hfill
    \begin{subfigure}{0.24\textwidth}
        \centering
        \includegraphics[width=\textwidth]{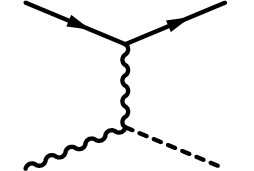}
    \end{subfigure}
    \hfill
    \begin{subfigure}{0.24\textwidth}
        \centering
        \hspace*{119.55139pt} 
    \end{subfigure}
    \hfill
    \begin{subfigure}{0.24\textwidth}
        \centering
        \hspace*{119.55139pt}
    \end{subfigure}
    \caption{$1 + 2 \rightarrow 3 + a$ processes in the KSVZ model. Wiggly lines represent gluons, arrowed lines represent fermions and dashed lines represent axions.}
    \label{fig:processes}
\end{figure}

The specificities of each axion model are encoded in the
various $c$ Wilson coefficients. As we mentioned, here we focus on thermal axion production in the KSVZ model~\cite{Kim:1979if,Shifman:1979if},
where the axion can only interact with gluons, implying $c_1=c_2=c_t=0$ and $c_3=1$.
In this model the only  processes  for producing one axion are, at leading 
order (LO) in the coupling $g_3$, the $2\leftrightarrow 2$ ones in Figure \ref{fig:processes}. 
There is just one free parameter, that is the axion scale $\fPQ$.
This model is therefore an ideal testing ground for the exploration of the reliability of perturbative 
determinations of the rate and it is at the same time of high phenomenological interest.
According to the analysis of \cite{DEramo:2021psx,DEramo:2021lgb}, values of $\fPQ$ close to the 
lower bound resulting from astrophysical constraints could give a $\Delta \neff$
contribution that would be accessible to the next-generation CMB telescopes.

In the remainder of this Section we first discuss the naive Boltzmann-equation approach to axion
production, its limitations and the emergence of collective effects in
Sec.~\ref{sub_naive}; this subsection is rather pedagogical, expert readers
can skip to Sec.~\ref{sub_resum}, where we discuss the implementation
of collective effects in the literature. Finally, in Sec.~\ref{sub_schemes}
we introduce our resummation schemes. Numerical results
are presented in the following Section.
A similar analysis has been recently
presented in~\cite{Becker:2023vwd} for
freeze-in dark matter production.

\subsection{The naive approach to the rate and the emergence of collective effects}
\label{sub_naive}
Let us now discuss the determination of the rate $\Gamma(k)$.
Naive kinetic theory and the Boltzmann equation  give
\begin{equation}
    \Gamma(k)^\text{naive}=\frac{1}{4k}\int\dd\Omega_{2\leftrightarrow 2}\sum\limits_{1,2,3\in \text{SM}}\left|\mathcal{M}_{1+2\rightarrow 3+a}\right|^2\frac{f_1(p_1)f_2(p_2)[1\pm f_3(k_1)]}{\nB(k)},\label{eq:prodrate}
\end{equation}
where considerations on the validity region of this approach will be presented soon.
 The equilibrium
phase-space distributions are $f_i(p)=\nB(p)$ when $i$ is bosonic and 
conversely $f_i(p)=\nF(p)\equiv(\exp(p/T)+1)^{-1}$ when fermionic.
$p_1,p_2,k_1,k$ are the momenta of the two incoming particles $1,2$ and of the outgoing $3$ and $a$ respectively, and the $1+\nB$ or $1-\nF$ accounts for final-state Bose enhancement and Pauli blocking. The factor of $1/4k$ corresponds to 
a standard $1/2k$ Lorentz-invariant phase space multiplying a factor of $1/2$ to account 
for identical particles in the initial state and to neutralize double-counting in the sums when 
particles 1 and 2 are different.

The final ingredient in \eqref{eq:prodrate}  are the matrix elements squared $|\mathcal{M}_{1+2\rightarrow 3+a}|^2$,
summed over all degeneracies of particles $1,2,3$. 
These are to be integrated over the $2\leftrightarrow 2$ phase space  $\dd\Omega_{2\leftrightarrow 2}$, as
given by Eq.~\eqref{phase_space} in App.~\ref{app_phase}.
We obtained said matrix elements using automated techniques.
The basic workflow we followed was as such,
as in~\cite{Ghiglieri:2020mhm}: first obtain the Feynman rules using \textsc{FeynRules}~\cite{Alloul:2013bka}, then call \textsc{FeynArts}~\cite{Hahn:2000kx} to generate all possible processes and finally use \textsc{FormCalc}~\cite{Hahn:2016ebn} to compute the amplitudes associated with each process and square them. 
The Lagrangian implemented in the model file for \textsc{FeynRules} was Eq.~\eqref{eq:laxion} with the
Wilson coefficients later set to the values of the KSVZ model.
Our results agree with those of~\cite{Graf:2010tv} and read
\begin{equation}
    |\mathcal{M}_{g+g\rightarrow g+a}|^2 = \frac{g_3^6(N_c^2-1)N_c}{32\pi^4\fPQ^2}\bigg[\frac{s u}{t}+\frac{ st}{u}+\frac{t u}{s}\bigg],\qquad
     |\mathcal{M}_{q+g\rightarrow q+a}|^2 = -\frac{g_3^6C_FN_c}{64\pi^4\fPQ^2}\frac{s^2+u^2}{t},
     \label{matrixelements}
\end{equation}
where the quark is taken to be a Dirac quark \jg{of negligible mass}
and the $q\bar{q}\to ga$ can be obtained by crossing. 
$C_F=(N_c^2-1)T_F/(N_c)$ is the quadratic Casimir of the fundamental representation of SU($N_c=3$)
and $T_F=1/2$. $s$, $t$
and $u$ are the usual Mandelstam \kb{invariants}, $s\equiv(P_1+P_2)^2$, $t\equiv(P_1-K_1)^2$, with 
on-shell external momenta $P_i=(p_i,\vec{p}_i)$. We denote four-momenta by capital letters,
moduli of three-momenta by lowercase ones. Our metric is the mostly-minus one.

Hence, one can see that Eq.~\eqref{eq:prodrate} is predicated on the assumption of the existence
of on-shell massless particles in the thermal plasma of the early universe.
In a thermal QCD plasma, particle states that are massless in vacuum develop \emph{thermal masses}
of the order of $g_3T$ \cite{Klimov:1981ka,Klimov:1982bv,Weldon:1982bn}.\footnote{This is true in a very broad 
class of quantum field theories, including gauge theories, where the mass is of the order of the gauge coupling 
$g$ times the temperature. Here we concentrate for simplicity on the QCD 
sector of the SM due to its direct coupling to axions in our chosen model.}
These however are not important at LO for the external states, as the phase-space integral
is dominated by regions where the external states have momenta of $\mathcal{O}(T)$.

The second so-far unstated assumption undergirding Eq.~\eqref{eq:prodrate} can be 
made explicit by plugging in Eq.~\eqref{eq:prodrate} the matrix elements in Eq.~\eqref{matrixelements}.
We find  
\begin{align}
    \Gamma(k)^\text{naive}_\text{KSVZ}=\frac{g_3^6(N_c^2-1)}{128\pi^4\fPQ^2k}\int\dd\Omega_{2\leftrightarrow 2}\bigg\{&
    N_c\bigg[2\frac{s u}{t}+\frac{t u}{s}\bigg]\frac{\nB(p_1)\nB(p_2)[1+ \nB(k_1)]}{\nB(k)}\nonumber \\
    &
    -2 T_F N_f\frac{s^2+u^2}{t}\frac{\nF(p_1)\nB(p_2)[1- \nF(k_1)]}{\nB(k)}\nonumber \\
    &+ T_F N_f\frac{t^2+u^2}{s}\frac{\nF(p_1)\nF(p_2)[1+ \nB(k_1)]}{\nB(k)}\bigg\},
    \label{eq:prodrate_full}
\end{align}
where $N_f$ denotes the number of quark flavors that can be considered light. We shall
return to this in Sec.~\ref{sec_prodrateres}. We have further used a relabeling 
of $P_1$ into $P_2$ and vice-versa to shuffle the $u$ denominators into $t$ ones
on the first and second line. These $t$ denominators are precisely linked 
to the assumption we need to make explicit: when integrating over the 
phase space, these denominators are responsible for would-be IR  divergences.
These $t$ denominators are best integrated using 
the $t$-channel parametrization detailed in Eq.~\eqref{tchannel} in App.~\ref{app_phase},
which keeps as explicit integration variables the frequency and momentum exchange $q^0$
and $q$, the momentum $p_1$ and an azimuthal angle $\phi$, with $P_2=K-Q$. $t$
then becomes $t=q_0^2-q^2$. At small $t$ we can then find the would-be divergent contribution
from a $t\ll s\sim T$ expansion, finding
\begin{align}
    \Gamma(k)^\text{naive div}_\text{KSVZ}=&\frac{g_3^6(N_c^2-1)}{2^{13}\pi^7\fPQ^2k^2} 
 \int_{q\ll k} \! {\rm d}q \int^q_{-q} \! {\rm d}q^{ }_0 
 \int_{0}^\infty \! {\rm d}p^{ }_1 
 \int_{-\pi}^{\pi} \! \frac{{\rm d}\varphi}{2\pi}\nonumber \\
 &\times\bigg\{
    -2N_c\frac{s^2}{t}\nB(p_1)[1+ \nB(p_1)]
    -4 T_F N_f\frac{s^2}{t}\nF(p_1)[1- \nF(p_1)]\bigg\}.
    \label{eq:prodrate_div}
\end{align}
The expression for $s$ in Eq.~\eqref{stchannel} can be used to simplify 
the matrix element squared into
\begin{equation}
    \frac{s^2}{t}\stackrel{k,p_1\gg Q}{\approx}\frac{4k^2p^{ 2}_1(q_0^2-q^2)
   (1 
  -\cos\varphi\,)^2}{q^4_{ }}.
  \label{divmatelem}
\end{equation}
Plugging this in Eq.~\eqref{eq:prodrate_div} and performing the $p_1$ \jg{and $\varphi$} integration\jg{s} leads to
\begin{align}
    \Gamma(k)^\text{naive div}_\text{KSVZ}=&\,\frac{g_3^6 (N_c^2-1)T^3}{2^{11}\pi^5\fPQ^2} 
 \int_{q\ll k} \! {\rm d}q \int^q_{-q} \! {\rm d}q^{ }_0 
\frac{(q^2-q_0^2)}{q^4_{ }}\bigg\{
    N_c
    +T_F N_f\bigg\}
    \nonumber\\
    =&\,\frac{g_3^4 (N_c^2-1)T \mD^2}{ 2^{9}\pi^5\fPQ^2} 
 \int_{q\ll k} \! 
\frac{{\rm d}q}{q},
    \label{eq:prodrate_div_final}
\end{align}
where we have noted that the first line is proportional to the QCD \emph{Debye mass}, which reads at LO
\begin{equation}
    \label{defmd}
    \mD^2=\frac{g_3^2 T^2}{3}\big(N_c+T_FN_f\big)\,.
\end{equation}
This quantity is the so-called chromoelectric screening mass: it is of order $g_3T$ and it
marks the first appearance of collective excitations, which screen long-wavelength electrostatic 
gluons. In our case it multiplies the would-be IR divergence $\dd q/q$; as soon as $q\sim g_3 T$
our naive description in terms of bare, in-vacuum matrix elements breaks down, precisely
because it cannot, by construction, capture collective effects arising at that scale. This is
precisely the second unstated assumption: the naive approach is only valid when $Q\gg g_3T$.

In our specific case, the naive approach misses \emph{Landau damping}, sometimes also called
\emph{dynamical screening} \cite{Weldon:1982aq,Baym:1990uj}. This collective effect ends up shielding the would-be divergence
thanks to the emergence not just of a screening mass but also of a width for space-like soft gluons,
corresponding to their damping rate. Proper accounting of this phenomenon requires then
abandoning, at least for the $Q\sim g_3 T$, $K\sim P_1\sim T$ region, the naive approach
based on bare matrix elements and replacing it with some form of resummation encoding
these collective effects.

This is of course not new. Indeed, shortly after the development of the Hard Thermal
Loop effective theory~\cite{Braaten:1989mz,Braaten:1991gm}, which describes 
in a gauge-invariant manner these collective excitations through resummed propagators
and vertices, \cite{Braaten:1991dd} applied HTL resummation to the abelian version of the
problem at hand.
In the next Subsection we explain in more detail this and other implementations 
of resummation in the literature and introduce ours.

We conclude this Subsection by noting one final limitation of the naive approach: it 
also requires the axion to be hard, with momentum obeying $k\gg g_3 T$. 
For $k\lesssim g_3T$ one can on 
the other hand expect collective effects to play an important, LO role, leading again 
to a breakdown of Eq.~\eqref{eq:prodrate}. Properly dealing with this region 
requires more intricate approaches than the relatively simple resummations presented in
the next Subsections; indeed, there currently exist no calculation of thermal production 
rates for light particles that are valid in the $k\sim g_3 T$ regime, neither in the context 
of early-universe cosmology --- see~\cite{Besak:2012qm,Ghiglieri:2020mhm} for ultrarelativistic
right-handed neutrinos and gravitational waves --- or in hot QCD --- 
see \cite{Arnold:2001ba,Arnold:2001ms,Ghiglieri:2013gia}. We will return 
later to this, when discussing extrapolating  to this regime.

\subsection{Collective effects and resummation in the literature}
\label{sub_resum}

As we mentioned, Landau damping was first introduced to the context of the axion
production rate in~\cite{Braaten:1991dd} for the abelian case.\footnote{%
\label{footnote_abelian}This approach
is based on HTL resummation. As the photon and gluon HTL propagators
differ only in the respective Debye masses, nothing changes qualitatively
in the non-abelian case at leading order.} It 
can be accounted for by applying Thermal Field Theory (TFT) and HTL resummation.
The former yields the well-known relation
\begin{equation}
    \Gamma(k)=\frac{1}{k}\mathrm{Im}\,\Pi^R_a(k,k)\,,
    \label{defgammapi}
\end{equation}
where $\Pi^R_a(k,k)$ is the retarded axion self-energy evaluated on the light cone, i.e.
\cite{Bodeker:2015exa}
\begin{equation}
    \label{defpi}
    \Pi^R_a(K)=\int \mathrm{d}^4 X e^{iK\cdot X}\theta(x^0)\langle [J(X),J(0)]\rangle\,,\quad\text{where}\quad
    J\stackrel{\rmii{KSVZ}}{=}\frac{\alpha_3}{8\pi\fPQ}G^b_{\mu\nu}\tilde{G}^{\mu\nu}_b\,.
\end{equation}
In arbitrary models $J$ is the operator coupling to the axion field, $\mathcal{L}\supset-aJ$.
With the methods
of~\cite{Bodeker:2015exa} one can show that Eq.~\eqref{defgammapi} is
correct to first order in the axion-SM couplings
and to all orders in the internal SM couplings. 

\begin{figure}[t]
    \centering
    \includegraphics[width=0.5\textwidth]{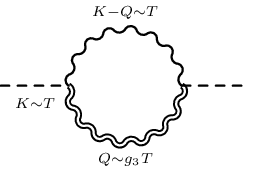}
      \includegraphics[width=0.4\textwidth]{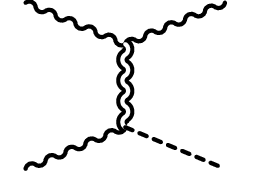}
    \caption{
    \jg{Left: }one-loop axion self-energy. The upper wiggly line denotes a bare gluon with a hard thermal momentum, while
    the double wiggly line stands for a soft, HTL-resummed gluon.
\jg{Cutting this one-loop axion self-energy corresponds to the square of
the diagram on the right. We don't show its analogue with a quark
or antiquark external scatterer replacing the top ingoing and outgoing gluons.}}
    \label{fig:axionself}
\end{figure}

In the approach of \cite{Braaten:1991dd} one can then use
Eq.~\eqref{eq:prodrate_full} up to an IR cutoff $q^*$ on the $\dd q$ integration.
This cutoff must be chosen such that $g_3 T\ll q^*\ll T$, so that 
HTL resummation can be performed only for $q\ll q^*$. There one can use 
Eq.~\eqref{defgammapi}. Evaluating Eq.~\eqref{defpi}
one encounters at LO for $k\gtrsim T$ and soft $Q$ the diagram \jg{on the left of} Fig.~\ref{fig:axionself},
where the 
lower 
gluon line carries the loop momentum $Q\sim g_3T$, while the upper line 
carries $K-Q\approx K\gtrsim T$. Hence it is only the lower
line that needs HTL resummation, whereas the upper one can be kept in its bare form. An identical contribution comes
from the diagram where the upper gluon is soft and lower 
one hard.

We recall that HTLs are the infrared, gauge-invariant
limits of one-loop thermal amplitudes with hard loop momentum $P\sim T$
and soft external momenta $Q_i\sim g_3 T$. HTL resummation accounts 
for the emergence of collective effects at that scale.
We also note that the imaginary part of the diagram \kb{on the left} in 
Fig.~\ref{fig:axionself} is given by its cut; 
the cut upper bare gluon is proportional to $\delta((K-Q)^2)\approx\delta(2Q\cdot K)$, which forces $Q$ to be space-like in the 
$K\gg Q$ limit. This in turn makes the cut HTL gluon  Landau-damped. Hence, the cut of this diagram corresponds to \jg{the square of the diagram shown on the right of Fig.~\ref{fig:axionself} and its counterpart with external quarks or antiquarks replacing the incoming and outgoing gluons at the top.}

The HTL-resummed result then yields \cite{Braaten:1991dd}
\begin{equation}
    \Gamma(k)_\text{KSVZ}^\text{HTL \cite{Braaten:1991dd}}=\frac{g_3^4 (N_c^2-1) T}{2^{10} \pi ^6 \fPQ^2} \int_0^{q^*} \dd q\, q^3\int_{-q}^q \frac{\dd q^0}{q^0}
    \left(1-\frac{q_0^2}{q^2}\right)\bigg[\rho_L(Q)+ \left(1-\frac{q_0^2}{q^2}\right)\rho_T(Q)\bigg]\,,
    \label{HTLfinal}
\end{equation}
where $\rho_L$ and $\rho_T$ are the longitudinal and transverse HTL spectral function,
obtained from the retarded propagators in Eqs.~\eqref{htllong}-\eqref{htltrans} 
from $\rho\equiv G_R-G_A$. \jg{ We remark that, up to the overall prefactor,
Eq.~\eqref{HTLfinal} agrees with the leading-order soft contribution 
to the ``transverse momentum broadening coefficient'' $\hat{q}$ in QCD plasmas ---
see for instance Eqs.~(36) and (59) in \cite{Ghiglieri:2015zma}. We shall make
use of this later.}

One can easily check from the UV limit $q\sim q^*\gg g_3 T$ of the spectral functions, i.e.
\begin{equation}
\label{UVspfs}
    \rho_L(Q)\to \frac{\pi \mD^2 q^0}{q^5}\,,\qquad \rho_T(Q)\to -\frac{\pi \mD^2 q^0}{2q^3 Q^2}\,,
\end{equation}
that Eq.~\eqref{HTLfinal} contains a UV logarithmic divergence in $q^*$ that cancels
the opposite IR one of the bare, naive contribution. Hence, summing the two gives rise to 
a finite, regulator-independent contribution. This is the approach that has been
implemented for the KSVZ model in \cite{Graf:2010tv},\footnote{%
\label{foot_mass}The earlier implementation in~\cite{Masso:2002np} 
cut off the IR divergence with the Debye mass and is thus 
only valid at leading-logarithmic accuracy.
} i.e.
\begin{equation}
    \Gamma(k)_\text{KSVZ}^\text{strict LO \cite{Graf:2010tv}}=\Gamma(k)^\text{naive}_\text{KSVZ}\bigg\vert_{q>q^*}+\Gamma(k)_\text{KSVZ}^\text{HTL \cite{Braaten:1991dd}}
 \label{strictLO}
\end{equation}

This implementation is a \emph{strict LO scheme} for hard axions with $k\gtrsim T$: 
it does not resum
any partial subset of higher order effects --- HTL resummation in the $Q\ll K\gtrsim T$
range is a strict LO effect. The resulting function of momentum \kb{contains} a 
$\ln(k/\mD)$ term, which clearly highlights the validity region 
of the calculation: when \emph{extrapolating} to $k<\mD$ this term becomes negative and for $k\ll \mD$
it turns the entire rate negative. This is easily seen in Fig.~\ref{fig:prodratetemp},
where the dotted green curve becomes negative at small $k$. As the 
temperature decreases the QCD coupling increases and the scale
separation between $k\gtrsim T$ and $\mD$ becomes blurry, making
the $k$ range with negative unphysical rates larger. 
This is a common feature to calculations
adopting the strict LO scheme defined by the method of~\cite{Braaten:1991dd} ---
see the detailed analysis in~\cite{Brandenburg:2004du} for the analogous case of thermal
axino production. 

As explained there, the optimal way to proceed would be to carry out a calculation
in the $k\sim g_3T$ regime. However, as we have mentioned, no such calculation
has been performed in the literature for any production rate of 
light-like particles. Here we do not present such an ambitious calculation; rather,
we introduce to the context of axion production two different 
schemes for the axion rate for $k\gtrsim T$ that extrapolate better
to the $k\lesssim \mD$ regime. This is presented 
in the next subsection. 

\jg{We conclude by commenting on} another approach in 
the literature~\cite{Salvio:2013iaa}, based on a gauge-dependent resummation \jg{first 
introduced for gravitinos in \cite{Rychkov:2007uq}.
It relies on the idea of using the full, rather than HTL, gluon spectral
functions at one loop in Feynman gauge, thus avoiding the need of separating 
the soft, HTL resummed region from the hard, bare one and giving 
rise to a positive-definite integrand and rate. 
While this might sound like a sound and sensible approach, 
it is well known, starting from \cite{Kalashnikov:1980tk,Kajantie:1982xx}, that 
the full one-loop transverse spectral function contains a gauge-dependent 
double-pole divergence for $q^0\approx 0$, $q\sim g_3^2T$, signaling the emergence
of an unphysical sensitivity to the chromomagnetic scale
where pertubation theory breaks down~\cite{Linde:1980ts}. 
\cite{DEramo:2012uzl} carried out a similar analysis 
for momentum broadening in QCD plasmas and concluded that 
resumming full self-energies for all $Q$ in Feynman gauge --- or any other gauge --- 
 gives rise to the same pathological divergence. As the soft contribution
to transverse momentum broadening is, up to prefactors, the same as in Eq.~\eqref{HTLfinal}, their conclusions apply to this case as well.
We shall return to this  }\jg{later in the main text. We will also}  \jg{prove the emergence
of this extra unphysical divergence in great detail in App.~\ref{app_salvio}.}

\subsection{Novel LO-accurate schemes }
\label{sub_schemes}

In this section we shall exploit the freedom to resum subsets of higher-order
terms in the $k\gtrsim T$ LO rate to define two new \emph{schemes}
that are equivalent to the strict LO one at vanishing coupling. \kb{This is 
a commonplace technique in hot QCD, where it has been used to study the intrinsic 
theory uncertainty of the perturbative expansion --- see App.~\ref{app_QCD} for a brief overview.}

Our first scheme is a slight rework of the strict LO; it was introduced for
right-handed-neutrino production in~\cite{Besak:2012qm,Ghiglieri:2016xye} and for 
 gravitational-wave production in~\cite{Ghiglieri:2020mhm}. In a nutshell,
one subtracts the divergent part found in Eq.~\eqref{eq:prodrate_div_final}
from the naive rate, leading to the finite expression
\begin{equation}
    \label{subtrrate}
    \Gamma(k)^\text{naive subtr}_\text{KSVZ}= \Gamma(k)^\text{naive}_\text{KSVZ}
    -\frac{3g_3^4 (N_c^2-1)T\mD^2}{2^{11}\pi^5\fPQ^2} 
 \int^k_{-\infty} \! {\rm d}q^{ }_0  \int_{|q^0|}^{2k-q^0} \! {\rm d}q
\frac{(q^2-q_0^2)}{q^4_{ }},
\end{equation}
As we show in detail in App.~\ref{app_phase}, we use the methods of 
\cite{Besak:2012qm} to carry out all integrals except the $q_0$
and $q$ analytically in $\Gamma(k)^\text{naive}_\text{KSVZ}$,
finding a finite expression after the subtraction above.
Note that we have kept the original integration limits for these variables, see Eq.~\eqref{tchannel}.
To complement Eq.~\eqref{subtrrate} with the necessary HTL resummation 
we slightly change Eq.~\eqref{HTLfinal} into
\begin{equation}
    \Gamma(k)_\text{KSVZ}^\text{HTL \cite{Ghiglieri:2016xye,Ghiglieri:2020mhm}}=\frac{g_3^4 (N_c^2-1) T}{2^{10} \pi ^6 \fPQ^2} \int^k_{-\infty} \! {\rm d}q^{ }_0  \int_{|q^0|}^{2k-q^0} \! {\rm d}q \frac{q^3}{q^0}
    \left(1-\frac{q_0^2}{q^2}\right)\bigg[\rho_L(Q)+ \left(1-\frac{q_0^2}{q^2}\right)\rho_T(Q)\bigg]\,,
    \label{HTLlimits}
\end{equation}
where we have kept the same integration limits of Eq.~\eqref{subtrrate}: since the UV limit of Eq.~\eqref{HTLlimits}
is subtracted there, this is now necessary.
As per \cite{Ghiglieri:2016xye,Ghiglieri:2020mhm} we can now 
change variables from $q$ to $q_\perp\equiv\sqrt{q^2-q_0^2}$.\footnote{%
\label{foot_qperp}This $q_\perp$ corresponds to the component of $\vec{q}$
orthogonal to $\vec{k}$ only at zeroth order in the $q\ll k$ expansion. In general 
one has $q_\perp^2=(q^2-q_0^2)((2k-q^0)^2-q^2)/(4k^2)$.}
Up to order-$g_3^2$ effects we can then simplify the integration region as
\begin{align}
    \Gamma(k)_\text{KSVZ}^\text{HTL \cite{Ghiglieri:2016xye,Ghiglieri:2020mhm}}=&\frac{g_3^4 (N_c^2-1) T}{2^{10} \pi ^6 \fPQ^2} \int^{\infty}_{-\infty} \! {\rm d}q^{ }_0  \int_{0}^{2k} \! {\rm d}q_\perp\, q_\perp \frac{q_\perp^2}{q^0}
     \bigg[\rho_L(Q)+\frac{q_\perp^2}{q^2}\rho_T(Q)\bigg]\nonumber\\
     =&\frac{g_3^4 (N_c^2-1) T}{2^{9} \pi ^5 \fPQ^2}  \int_{0}^{2k} \! {\rm d}q_\perp\, q_\perp^3
     \bigg[\frac{1}{q_\perp^2}-\frac{1}{q_\perp^2+\mD^2}\bigg]\nonumber\\
     =&\frac{g_3^4 (N_c^2-1) T\mD^2}{2^{10} \pi ^5 \fPQ^2}\ln\bigg(1+\frac{4k^2}{\mD^2}\bigg)\,,
    \label{HTLlimitsfinal}
\end{align}
where in going from the first to the second-line we used a sum rule 
derived \jg{for $\hat{q}$} in \cite{Aurenche:2002pd,CaronHuot:2008ni} --- see also
\cite{Ghiglieri:2015zma,Ghiglieri:2020dpq} for more pedagogical
illustrations. Adding together Eqs.~\eqref{subtrrate}
and \eqref{HTLlimitsfinal} we then arrive at the definition
of the \emph{subtraction scheme}, namely
\begin{equation}
    \label{subtrratefinal}
     \Gamma(k)^\text{subtr}_\text{KSVZ}=\Gamma(k)^\text{naive subtr}_\text{KSVZ}+\Gamma(k)_\text{KSVZ}^\text{HTL \cite{Ghiglieri:2016xye,Ghiglieri:2020mhm}}\,.
\end{equation}
We refer to Eq.~\eqref{eq:prodratetotintsubtr} for details on our implementation.
It is important to note that $\Gamma(k)^\text{subtr}_\text{KSVZ}$,
as it is obtained from a simple rework of the gauge-invariant strict LO scheme, is 
gauge invariant. Furthermore,
$\Gamma(k)^\text{subtr}_\text{KSVZ}$ is no longer a strict LO scheme: indeed, Eq.~\eqref{HTLlimitsfinal}
contains higher-order terms for $k\gtrsim T$, where $\ln(1+4k^2/\mD^2)=\ln(4k^2/\mD^2)+\mD^2/(4k^2)+\ldots$.
Hence, these resummed terms start at relative order $g_3^2$. We have furthermore checked
that, upon replacing $\ln(1+4k^2/\mD^2)$ with $\ln(4k^2/\mD^2)$ we find  agreement
with the strict LO results of \cite{Graf:2010tv}, obtained by numerically integrating their expressions. Incidentally,
we remark that our implementation of the strict LO rate, as per Eq.~\eqref{eq:prodratetotintstrict}, converges faster. 

It is worth noting that $\ln(1+4k^2/\mD^2)$ is better behaved than $\ln(4k^2/\mD^2)$ when \emph{extrapolating} to 
$k\ll \mD$, as it becomes small and positive rather than large and negative. However, 
as Fig.~\ref{fig:prodratetemp} makes clear --- compare the dotted green line with the dashed blue one ---
this subtraction scheme is still not positive definite at all $k$. This can be understood from
having subtracted the divergent infrared part in the bare limit in Eq.~\eqref{subtrrate}
and having replaced it with its resummed version in Eq.~\eqref{HTLlimitsfinal}. The latter,
though positive definite, is smaller, thus opening the door to negativity outside of the 
validity region of this approach.

It is for this reason that we now introduce a positive-definite scheme. One possibility
would be to follow the method introduced in \cite{Arnold:2002zm,Arnold:2003zc}
for the effective kinetic theory of QCD. The authors exploited the liberty of defining rates 
that are equivalent to the strict LO at vanishing coupling but that differ by higher-order
terms. In detail, they singled out divergent channels/diagrams
and replaced the bare propagators therein with resummed, retarded HTL ones.
In our case, we would need 
to do so in the second and third diagram on the first line in Fig.~\ref{fig:processes}
and in the second diagram on the second line. Upon taking the square modulus of the 
amplitudes, as in Eq.~\eqref{eq:prodrate}, one naturally has a positive-definite quantity for all $k$.
This approach would have the drawback of requiring more intricate numerical integrations 
over the HTL propagators. The availability in the past decade of analytical results 
for the soft region, as in Eq.~\eqref{HTLlimitsfinal}, allows for the 
introduction of a numerically simpler scheme, first devised in~\cite{York:2014wja} for the 
numerical implementation of the kinetic theory in \cite{Arnold:2002zm}. Namely, we define our 
\emph{tuned mass scheme} as
\begin{align}
     \Gamma(k)^\text{tuned}_\text{KSVZ}=\Gamma(k)^\text{naive}_\text{KSVZ}\quad\text{with}&\quad
     \frac{s^2+u^2}{t}\to \frac{q^4}{2(q^2+\xi^2 \mD^2)^2}\frac{(s-u)^2}{t}+\frac{t}{2}\nonumber\\
     \text{and}&\quad
     2\frac{su}{t}\to -\frac{q^4}{2(q^2+\xi^2 \mD^2)^2}\frac{(s-u)^2}{t}+\frac{t}{2}\,,
         \label{deftuned}
\end{align}
where $\xi>0$ is an $\mathcal{O}(1)$ constant that must be \emph{tuned} so that this method reproduces
the analytical result given by Eq.~\eqref{HTLlimitsfinal} in the soft region for $g_3\ll 1$ and $k\gtrsim T$. In App.~\ref{app_phase} we give the full expression for $\Gamma(k)^\text{tuned}_\text{KSVZ}$
in Eq.~\eqref{eq:prodratetotinttuned} and we show explicitly that\footnote{
This scheme was recently applied in~\cite{Boguslavski:2023waw}
to the transverse momentum broadening coefficient of QCD. Our
result for $\xi$ agrees with theirs.}
\begin{equation}
    \label{defxi}
    \xi = \frac{e^{1/3}}{2}\,.
\end{equation}
This scheme too is gauge invariant, as it is obtained by suitably modifying
the $t-$channel denominators of gauge-independent
matrix elements. Furthermore,
as Eq.~\eqref{deftuned} makes clear, for $q\gg \mD$ 
this scheme resums higher-order effects thanks to 
the denominator $(q^2+\xi^2\mD^2)^2$. 
The  energy dependence of the statistical factors for incoming and outgoing particle states 
is left unchanged, differently from what happens in the soft
region of the strict LO and subtraction schemes --- see Eq.~\eqref{eq:prodrate_div}.
Hence this schemes does not assume $T\gg \mD$, even though it reduces to the
value obtained there.
We also remark that, differently from what one would naively expect from the expansion
of $q^4/(q^2+\xi^2\mD^2)^2$ for $q\sim T\gg \mD$,
the difference between the strict LO scheme and this one is not 
of order $g_3^2$ but rather of order $g_3$ when $k\gtrsim T$. This $\mathcal{O}(g_3)$
difference arises from the region where $Q\sim g_3T$ and \jg{$p_1$ is a gluon's momentum with} $p_1\sim g_3T$, see
\cite{CaronHuot:2007gq,Caron-Huot:2008dyw,Fu:2021jhl} for  analogous cases. 
\jg{This $\mathcal{O}(g_3)$ deviation emerges clearly from Fig.~\ref{fig:diff}, as we shall show in the next section.} \kb{We shall return to the topic of $\mathcal{O}(g_3)$ corrections later.}

Compared with the strict LO and subtraction schemes, the current one has a drawback,
namely that the dependence on the Debye mass is not factored out in an analytical part
such as Eq.~\eqref{HTLlimitsfinal}, which is readily evaluated. In this case 
one has two perform a 2D numerical integration for each value of $k/T$ and for each value of 
$\mD/T$; in App.~\ref{app_phase} we give details on the numerical implementation.
Results are shown in Fig.~\ref{fig:prodratetemp} as a solid red line. As expected, 
it is everywhere positive  and larger than the two other curves. As the difference 
between this rate and the strict LO and subtraction schemes is of order $g_3$, it remains
quantitatively distinct from these other two schemes for $k\gtrsim T$ even at the largest temperature considered there.

We wish to emphasize again that no scheme is leading-order accurate for $k\lesssim \mD$\jg{, whereas for $k\gtrsim T$ they all are, with the tuned one larger than the other two by an $\mathcal{O}(g_3)$ amount.}
In the following we shall consider the spread between the rates and the ensuing axion abundances
as a handle on the theory uncertainty of these leading-order calculations.
\jg{We think that this uncertainty could be better quantified by a determination 
of the NLO $\mathcal{O}(g_3)$ correction to the rate for $k\gtrsim T$,
along the lines of what has been carried out in \cite{CaronHuot:2008ni}
for transverse momentum broadening and in \cite{Ghiglieri:2013gia,Ghiglieri:2018dib}
for the photon production rate from a QCD plasma and its shear viscosity, 
respectively.} \kb{As we briefly review in App.~\ref{app_QCD}, these $\mathcal{O}(g_3)$ corrections
arising from the soft $g_3T$ momentum scale are the root cause of the oftentimes poor convergence
shown in perturbative calculations.}

\kb{In light of this, we think that a conservative estimate of their possible size would in turn give a 
conservative handle on the effect they would have. In~Eq.~\eqref{largeNLO} we thus exploit the direct proportionality between
the soft contribution to $\Gamma(k)_\text{KSVZ}$ and $\hat{q}$, as pointed out after Eq.~\eqref{HTLfinal}, to include the NLO corrections to the latter as a proxy for the size of potentially large NLO corrections 
to  $\Gamma(k\gtrsim T)_\text{KSVZ}$. We remark that, due to this proportionality,
these corrections will certainly be part of the full set of NLO corrections,
whose complete determination we leave to future work.
Let us furthermore note that these NLO $\mathcal{O}(g_3)$ corrections 
cannot be correctly captured by the approach of \cite{Rychkov:2007uq,Salvio:2013iaa}: these corrections require 
proper handling of soft external states and of vertices with soft external momenta. Conversely, the vertices and propagators in their resummed, Feynman--gauge self-energy are bare and can only describe hard quasi-particles.
}

\kb{We conclude by noting that}, once $k\lesssim g_3^4T$, the quasiparticle approach implicit in Eq.~\eqref{eq:prodrate}
breaks down completely and one enters the so-called hydrodynamic regime. There the 
axion can no longer resolve particles or even HTL quasiparticles but only longer-wavelength 
hydrodynamical-like excitations, which are encoded in transport coefficients, i.e. 
the zero-frequency slope of the spectral function of the underlying conserved charges. In
the case of photon production the charge would be the electromagnetic one 
and the transport coefficient the electrical conductivity, in the case 
of gravitational waves one has the energy-momentum tensor and the shear viscosity. In this
case one has the topological charge and the \emph{sphaleron rate},
which is proportional to the thermal axion production rate at vanishing momentum~\cite{McLerran:1990de}. 

Axions at such soft momenta give a very small contribution to $\Delta\neff$ --- see Eqs.~\eqref{eq:deltafirst}
and \eqref{def_endensity} --- and once $k\sim m_a$ they would not be
dark radiation. It has however recently been proposed in~\cite{Notari:2022zxo,Bianchini:2023ubu} to use non-perturbative
determinations of the strong sphaleron rate up to $k\gtrsim T$. Let us point out that 
a  recent parametrisation~\cite{Laine:2022ytc} of the spectral function of the topological charge 
finds a ``transport dip'' rather than a transport peak at small frequency $\omega$,
pointing to a small contribution of the $\omega\lesssim g_3^4T$ region sensitive to sphaleron dynamics.
It would be interesting to carry out a similar analysis for the spectral function at finite momentum,
which is the relevant one for on-shell axion production.

\section{Numerical results for the production rate}
\label{sec_prodrateres}

In this section we present our numerical results for the axion rate in the three schemes defined in
the previous section.
We refer to App.~\ref{app_phase} for details on the reduction to two-dimensional
numerical integrals.
We use the results of \cite{Laine:2019uua} for the QCD running coupling,
which are five-loop accurate and account for quark mass thresholds. With a renormalisation
scale $\mu=2\pi T$ this yields $\alpha_s(T=300\text{ MeV})\approx0.31$ and $\alpha_s(T=10\text{ TeV})\approx 0.06$.

As per the previous section, the rate depends on $N_f$, the number of light quarks.
To account for the effect of the electroweak phase transition giving mass to the heaviest quarks and subsequently leading to them dropping out of thermal equilibrium, we used a dynamical number of quarks flavors
\begin{equation}
    N_f(T)=3+\sum\limits_{i=c,b,t}\frac{\chi_i(T)}{\chi(m=0)},
    \label{defnfT}
\end{equation}
where
\begin{equation}
    \chi_i(T)=\int\frac{\dd^3\mathbf{p}}{(2\pi)^3}(-2\nF'(E_i))
\end{equation}
is the susceptibility of the quark flavor $i$ with energy $E_i\equiv\sqrt{k^2+m^2_i(T)}$. There $m_i(T)$ is the mass of quark $i$
 and $\chi(m=0)= T^2/6$.
We  use the results of \cite{Laine:2019uua} for the quark masses $m_i(T)$, which are evolved
to five loops in the QCD coupling and which account for the temperature dependence of the
Higgs expectation value.

\begin{figure}[t]
    \centering
    \begin{subfigure}[b]{0.45\textwidth}
        \centering
        \includegraphics[width=\textwidth]{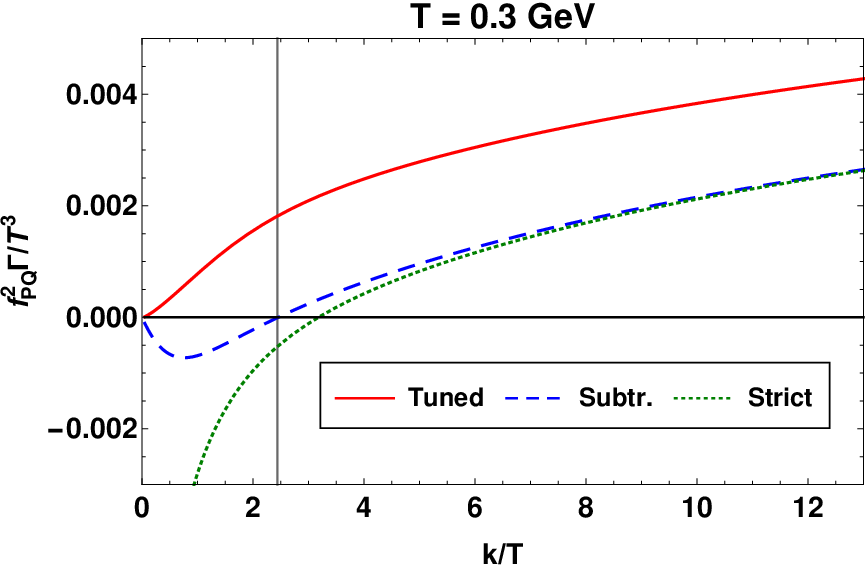}
        \caption{For $T=300$ MeV}
        \label{fig:prodratetemp1}
    \end{subfigure}
    \begin{subfigure}[b]{0.45\textwidth}
        \centering
        \includegraphics[width=\textwidth]{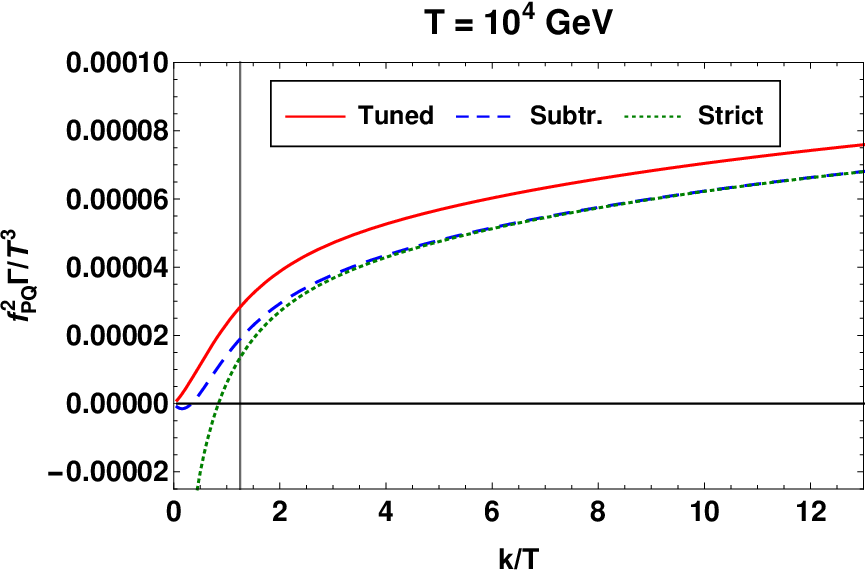}
        \caption{For $T=10$ TeV}
        \label{fig:prodratetemp300}
    \end{subfigure}  
    \caption{Production rate for all schemes at fixed $T$ as a function of the axion momentum $k$.
    The strict LO (``Strict''), subtraction (``Subtr.'') and tuned mass schemes (``Tuned'') are defined in Eqs.~\eqref{strictLO},
    \eqref{subtrratefinal} and \eqref{deftuned} respectively. See the main text for the choice of couplings 
    and quark mass thresholds. The vertical line corresponds to $k=\mD$.} 
    \label{fig:prodratetemp}
\end{figure}
In Fig.~\ref{fig:prodratetemp} we plot the rate $\Gamma(k)$ as a function of momentum
at a temperature close to the QCD transition --- $T=300$ MeV --- and at one above the electroweak 
transition, $T=10$ TeV. In the latter case the smaller coupling implies $\mD/T\approx 1$,
indicated by the vertical line for $k=\mD$. For $k>\mD$ we see that the three schemes 
are close, and that they respect our expectations: the strict LO and subtraction 
are much closer to each other than to the tuned mass scheme. Indeed, as we explained, for
$k\gtrsim T$ the strict LO and subtraction scheme differ by $\mathcal{O}(g_3^2)$, while
the tuned mass one is $\mathcal{O}(g_3)$ apart from them. 
At  $T=300$ MeV the correspondingly larger coupling shows well the effect of the extrapolation
to larger couplings and to $k\lesssim \mD$. In this case the $k$ range giving rise to
negative rates is much larger and the spread between the tuned mass and the other schemes more pronounced.

\begin{figure}[t]
    \centering
    \includegraphics[width=\linewidth]{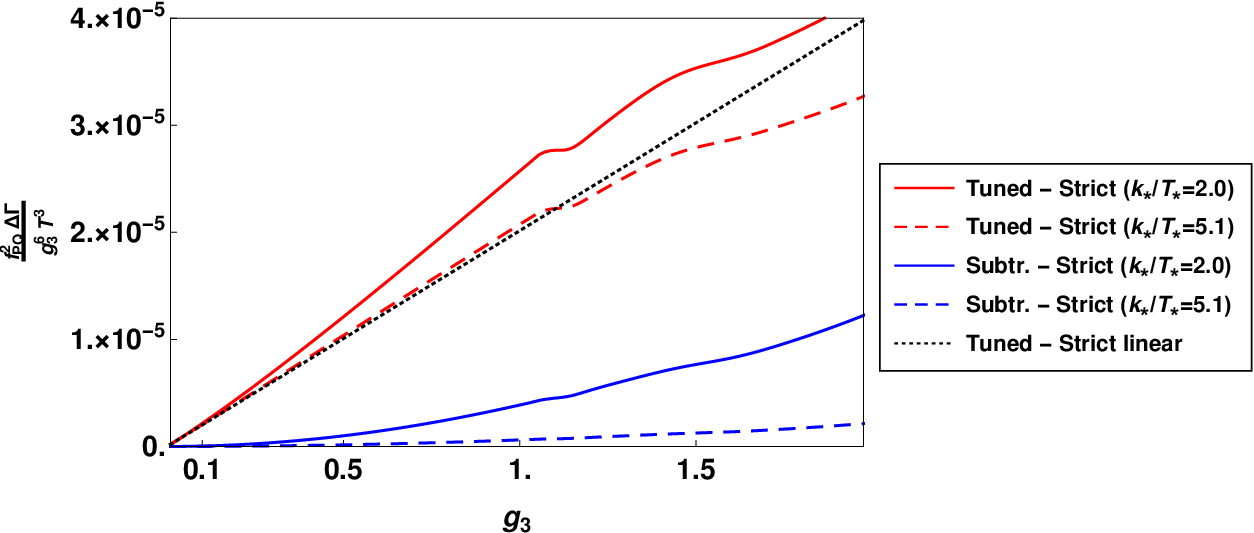}
    \caption{\jg{The difference between the rate in different schemes as a function of the coupling $g_3$. We normalise the vertical axis     by the leading power of the coupling, $g_3^6$, and by the dimensionful factor 
    $T^3/\fPQ^2$. We choose
    two momenta for which $k\gtrsim T$. See the main text for the definition
    of the `Tuned - Strict linear'' curve.
    The step-like features stem from changes in $N_f(T)$ through Eq.~\eqref{defnfT}.}
    }
    \label{fig:diff}
\end{figure}
\jg{To further highlight the deviations between the three schemes, 
in Fig.~\ref{fig:diff} we plot their difference normalised by the leading power of the coupling, $g_3^6$, and by the dimensionful factor 
    $T^3/\fPQ^2$. The plot then clearly shows  how the tuned scheme
    differs from the other two by an $\mathcal{O}(g_3)$ amount. 
    This is made particularly evident by the ``Tuned - Strict linear'' curve
    in dotted black. It shows the leading, $\mathcal{O}(g_3)$ contribution 
    to the difference between the tuned and strict schemes. It is computed 
    analytically in App.~\ref{app_phase}, where it is given by Eq.~\eqref{ogdiff}.
    It is momentum-independent.}

\begin{figure}[t]
    \centering
    \begin{subfigure}[b]{0.45\textwidth}
        \centering
        \includegraphics[width=\textwidth]{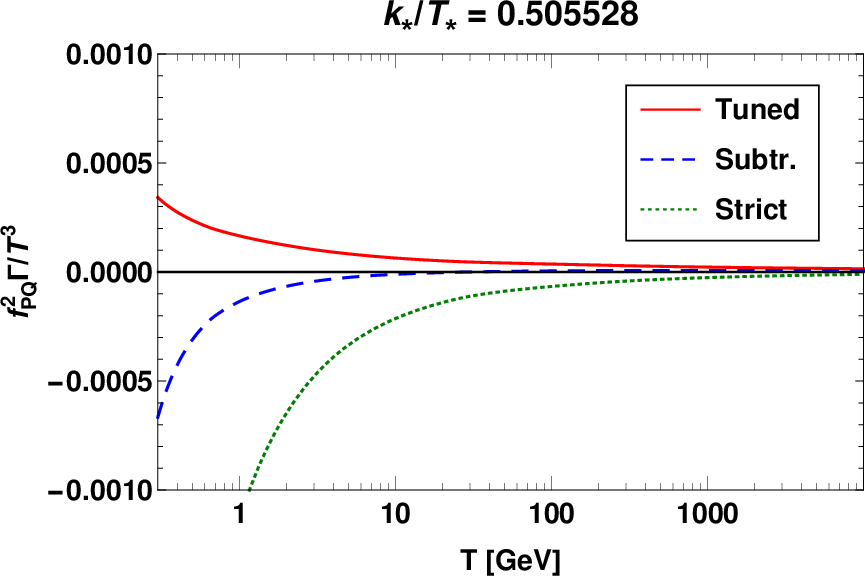}
        \caption{$8^{\mathrm{th}}$ mode}
        \label{fig:prodratemode8}
    \end{subfigure}
    \hfill
    \begin{subfigure}[b]{0.45\textwidth}
        \centering
        \includegraphics[width=\textwidth]{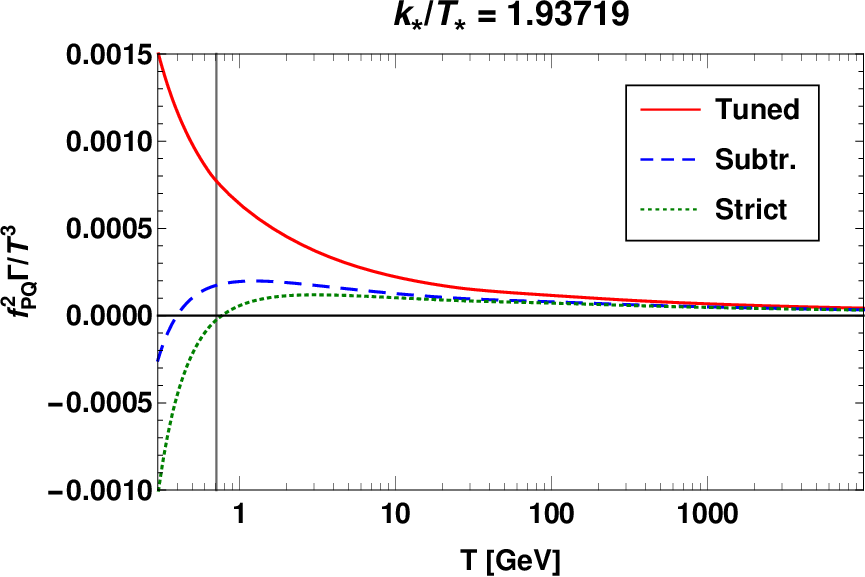}
        \caption{$30^{\mathrm{th}}$ mode}
        \label{fig:prodratemode30}
    \end{subfigure}

    \bigskip
    \begin{subfigure}[b]{0.45\textwidth}
        \centering
        \includegraphics[width=\textwidth]{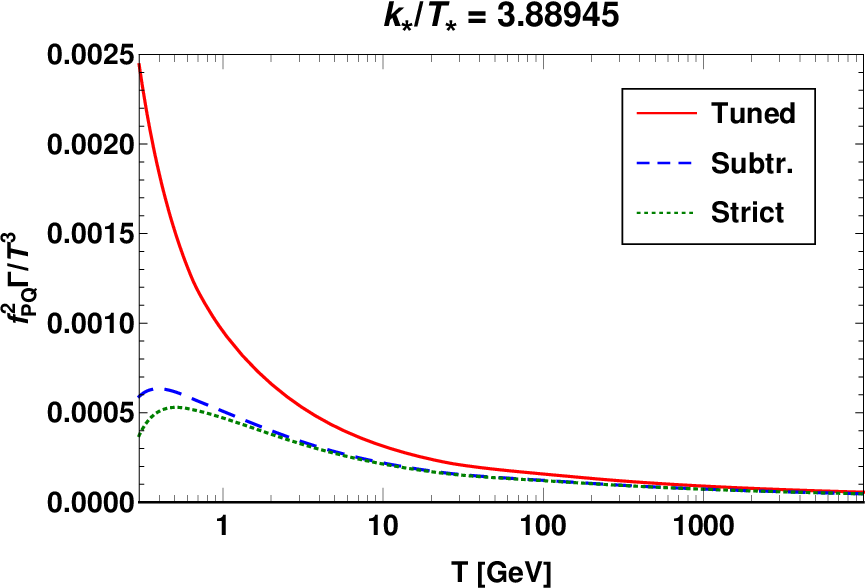}
        \caption{$60^{\mathrm{th}}$ mode}
        \label{fig:prodratemode60}
    \end{subfigure}
    \hfill
    \begin{subfigure}[b]{0.45\textwidth}
        \centering
        \includegraphics[width=\textwidth]{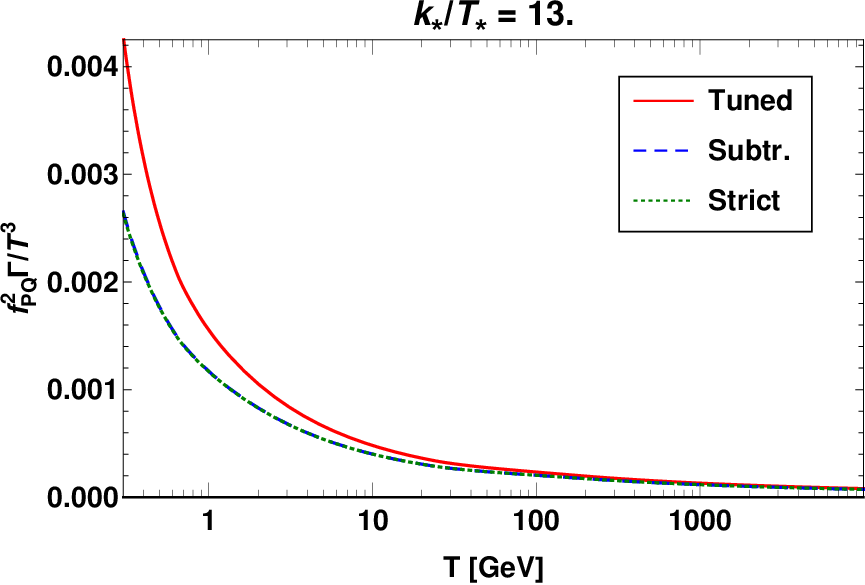}
        \caption{$200^{\mathrm{th}}$ mode}
        \label{fig:prodratemode200}
    \end{subfigure}    
    \caption{Production rate for all schemes at fixed $k_{*}$ as a function of the temperature, for a few
    test modes. The vertical line corresponds to the temperature for which $k(T)=\mD(T)$.
    In Fig.~\ref{fig:prodratemode8} this happens for $T>\tmax$, whereas in Figs.~\ref{fig:prodratemode60}--\ref{fig:prodratemode200} 
    this happens for $T<\tmin$.}
    \label{fig:prodratemode}
\end{figure}

In preparation for the solution of the rate equation, Eq.~\eqref{defintrate}, we proceed 
to tabulate the rates on a grid of temperature and comoving momentum. 
For the former we pick $300$ logarithmically spaced points between $\tmin=300\;\mathrm{MeV}$ 
and $\tmax=10\;\mathrm{TeV}$, with $200$ values below the electroweak 
crossover $T_\mathrm{EW}=160\;\mathrm{GeV}$ and $100$ values above.
We  define the (normalized) comoving momenta at a reference temperature $T_{*}=\tmin$ as $200$ evenly spaced points between $\kmin/T_{*}=0.05$ and $\kmax/T_{*}=13$.
For every other temperature, the momenta are blue-shifted using the relation
\begin{equation}
    \frac{k(T)}{T}=\frac{k_{*}}{T_{*}}\left(\frac{s(T)/T^3}{s(T_{*})/T_{*}^3}\right)^{1/3},
    \label{def_comoving}
\end{equation}
where  $k_{*}$ is the value of $k$ at $T_{*}$ and $s(T)$ is the SM entropy density, tabulated in \cite{Laine:2015kra}
accounting for interactions and quark mass thresholds. 

Numerical results obtained for the production rate, as well as for other quantities of phenomenological interest defined in Section \ref{sec_momdep}, have been made available on ZENODO \cite{ghiglieri_2024_10926565}.

In Fig.~\ref{fig:prodratemode} we now present some results for the production rate with respect to $T$ at fixed $k_{*}$.
This figure confirms the finding of Fig.~\ref{fig:prodratetemp}, namely that in the region of validity of the
calculations, i.e. $k\gtrsim T$, the strict LO and subtraction schemes agree perfectly.
On the other hand, when $k\lesssim \mD$ or, in the low $T$, high $g_3$ region, $k\sim T\sim \mD$, 
both methods yield unphysical negative rates.
However, it can be seen from Figures \ref{fig:prodratetemp300} and \ref{fig:prodratemode30} for example that 
the region where the subtraction production rate is negative is smaller than the region where the strict LO rate is negative.
The subtraction method also manages to keep the negative values at a more acceptable level, 
see \textit{e.g.} Figure \ref{fig:prodratetemp300} where the negative dip around $k=0$ is barely visible.

\kb{Next}, we compute and plot the momentum average $\langle \Gamma \rangle$ of the rate,
defined as
\begin{equation}
    \langle\cdots\rangle\equiv\frac{\int\frac{\dd^3 \vec{k}}{(2\pi)^3}\nB(k)(\cdots)}{\int\frac{\dd^3 \vec{k}}{(2\pi)^3}\nB(k)}.
    \label{defavg}
\end{equation}
In the interest of comparisons with previous works, we
recast it 
in the form of $F_3(T)$ in~\cite{Salvio:2013iaa,DEramo:2021lgb}.
Adapting it to our notation, it is expressed as:\footnote{Readers may have noticed that \eqref{eq:controlfunc} 
is lacking the 1PI effective coefficient $\tilde{c}^{\Psi}_g(T)$ present in \cite{DEramo:2021lgb}. 
This is because we consider the situation $T \ll m_{\Psi}$, with $m_\Psi$ the mass of the heavy KSVZ fermion. In 
this regime $\tilde{c}^{\Psi}_g(T)\to 1$.
}
\begin{equation}
    F_3(T)=\frac{512 \pi^5 \fPQ^2 \,\langle \Gamma\rangle }{ (N_c^2-1) g_3^4(T) T^3} .
    \label{eq:controlfunc}
\end{equation}
\begin{figure}[t]
    \centering
    \begin{subfigure}[b]{0.48\textwidth}
        \centering
        \includegraphics[width=\textwidth]{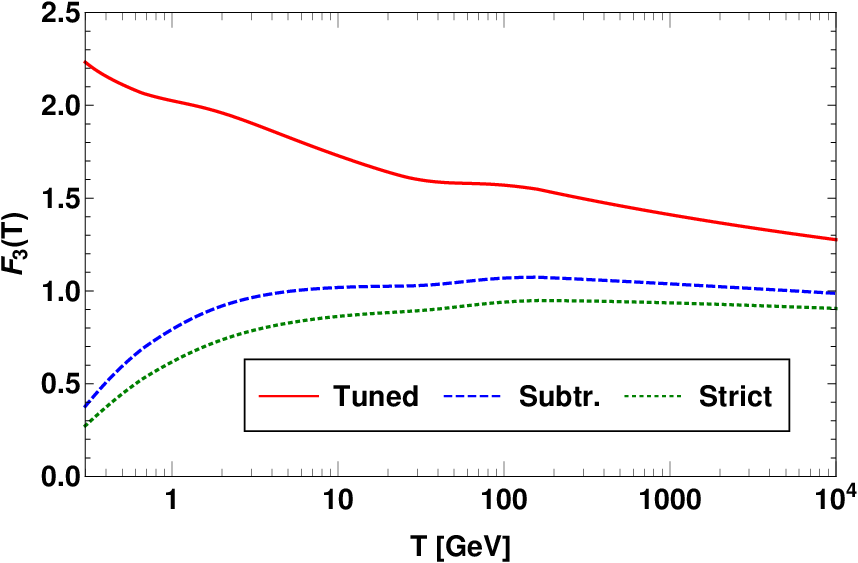}
        \caption{$F_3$ as a function of $T$}
        \label{fig:controlfunctemp}
    \end{subfigure}
    \begin{subfigure}[b]{0.48\textwidth}
        \centering
        \includegraphics[width=\textwidth]{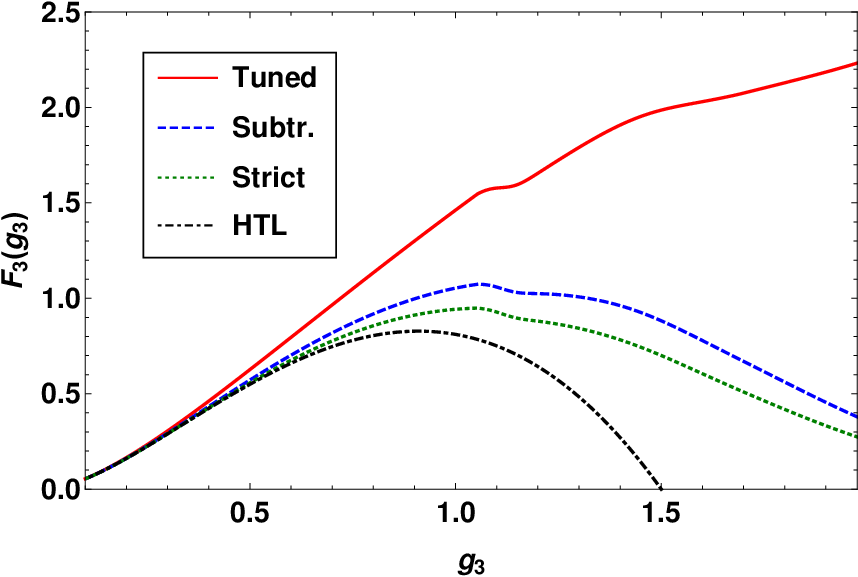}
        \caption{$F_3$ as a function of $g_3$}
        \label{fig:controlfuncg3}
    \end{subfigure}  
    \caption{Momentum-averaged rate $F_3$ for all three computational schemes. See the main text for 
    the definition of $F_3$ and for more remarks on the ``HTL'' curve in Figure \ref{fig:controlfuncg3}.
    The step-like features in the two plots stem from changes in $N_f(T)$ through Eq.~\eqref{defnfT}.}
    \label{fig:controlfunc}
\end{figure}

Numerical results for $F_3$ for all considered computation schemes 
are plotted in Fig.~\ref{fig:controlfunc}.
A comparison with the so-called ``HTL'' rate is also provided.
This rate is defined in \cite{Salvio:2013iaa} as
\begin{equation}
    F_3^\mathrm{HTL}=g_3^2\ln\left(\frac{1.5^2}{g_3^2}\right).
    \label{eq:htlcontrolfunc}
\end{equation}
It corresponds to the strict LO scheme of \cite{Graf:2010tv} with constant $N_f=6$. 
One important thing to note for the integral in \eqref{defavg} is that, as we shall
motivate in the next Section, we exclude momentum regions giving rise to unphysical negative rates from the integration range. 
This is different from what is done in \cite{Graf:2010tv} --- see also footnote~\ref{foot_neg}.
This is why $F_3^\mathrm{HTL}$ becomes negative for $g_3\gtrsim 1.5$.
Conversely, in the weak-coupling limit, $g_3\to 0$, we have $N_f(T)\to 6$ 
and we are also in the region of validity of the strict LO approach, so $\langle \Gamma\rangle$ remains positive.
We therefore expect our results to converge towards \eqref{eq:htlcontrolfunc}.
As we can see on Figure \ref{fig:controlfuncg3} this is indeed what happens. One also sees clearly
once more the difference between the $\mathcal{O}(g_3^2)$ spread of the subtraction and strict LO curves
and their $\mathcal{O}(g_3)$ difference with the tuned mass one.

\kb{Moreover}, we remark that the values of $F_3$ resulting from the \jg{gauge-dependent} scheme of \cite{Salvio:2013iaa} \jg{with its aforementioned pathologies} are
significantly larger. \cite{Salvio:2013iaa} finds $F_3\approx 5$ for $g_3\approx 1.1$, which is 3
to 5 times larger than our results. The numerical implementation of this scheme
in~\cite{DEramo:2021lgb} results in $F_3(T=10^4\text{ GeV})\approx 3$ and $F_3(T=2\text{ GeV})\approx 16.5$. Our values
for the tuned mass scheme are a factor of 2 lower at $T=10^4\text{ GeV}$ and a factor of 8 lower at  $T=2\text{ GeV}$.
\jg{As we mentioned}, in App.~\ref{app_salvio} we argue that the gauge-dependent resummation performed in~\cite{Salvio:2013iaa}
introduces an unphysical, gauge-dependent sensitivity to the 
chromomagnetic non-perturbative scale $g_3^2 T$ \cite{Linde:1980ts}.
\jg{This sensitivity shows up as}
 an extra \jg{double-pole} divergence, whose numerical handling
in \cite{Salvio:2013iaa,DEramo:2021lgb} is at the moment unclear.
\jg{Our position is that these results are not to be trusted,
as they most likely rely on some form of numerical regularisation
of a genuine, non-integrable divergence.}

\kb{Finally, let us give a first assessment of the possible size of NLO corrections.
As we explained at the end of the previous section, in Eq.~\eqref{largeNLO} 
we just add the very large 
$\mathcal{O}(g_3)$ corrections to $\hat{q}$ to the strict LO rate, to act as a proxy
for the expected large NLO corrections. As that equation shows, this $\mathcal{O}(g_3)$
contributions amounts to a positive,
momentum-independent factor multiplying $g_3^7T^3/\fPQ^2$, which thus shift the rate
upwards for all momenta. We find that for $T=10^4\text{ GeV}$ the value of $F_3$ 
from our NLO Ansatz is twice that from the tuned scheme. At our lowest
value, $T=300\text{ MeV}$, the NLO Ansatz $F_3$ is almost 9 times larger than its 
counterpart from the tuned scheme. }

\kb{As expected, our Ansatz results in very large 
corrections, which overtake any LO-equivalent scheme even at our highest temperature,
reproducing the behaviour already observed in thermal QCD calculations~\cite{CaronHuot:2008ni}. In the next section we will see the effect
that these potential corrections would have on $\Delta \neff$. Here we remark that
it has been suggested, in the hot QCD literature, that the apparent poor convergence may arise 
from an overscreeened LO rather than from
large NLO corrections \cite{Muller:2021wri,Ghiglieri:2022sui} --- see App.~\ref{app_QCD}
for more details. Efforts are currently underway in~\cite{Panero:2013pla,Moore:2019lgw} to 
determine the soft contribution to $\hat{q}$ --- and hence to the axion rate --- non-perturbatively
to all orders. These first results suggest that the non-perturbative value
should be smaller than the large, perturbative NLO one. 
We plan to return to this issue in a future publication.}

\section{Momentum-dependent axion freeze-out above the QCD crossover}
\label{sec_momdep}

We now turn to the evaluation of the phenomenological quantity of interest $\Delta\neff$, which measures the impact of thermal axion production on the total energy density of the Universe.
It is defined as
\begin{equation}
    \Delta\neff=\frac{8}{7}\left(\frac{11}{4}\right)^{4/3}\frac{e_a(T_\mathrm{CMB})}{e_{\gamma}(T_\mathrm{CMB})},\label{eq:deltafirst}
\end{equation}
where $e_a$ is the axion energy density, $e_{\gamma}$ is the photon energy density (from the Cosmic Microwave Background) and $T_\mathrm{CMB}\approx 0.3\;\mathrm{eV}$ is the decoupling temperature of the CMB.
The number and energy densities of a (massless) particle $X$ are defined as:
\begin{align}
    \label{def_ndensity}
    &n_X=g_X\int\frac{\dd^3\vec{k}}{(2\pi)^3}\,f_X(k),
    \\
    \label{def_endensity}
    &e_X=g_X\int\frac{\dd^3\vec{k}}{(2\pi)^3}\,k\,f_X(k),
\end{align}
where $f_X$ is the phase-space distribution of $X$ and $g_X$ its degeneracy.

The most natural way to obtain $\Delta \neff$ is to solve the momentum-dependent
Boltzmann equation, Eq.~\eqref{defintrate}, which we can rewrite as
\begin{equation}
        \dv[f_a(k(T))][\ln(T_*/T)]=\frac{\Gamma(k(T))}{3 H(T) c_s^2(T)}\,\left[\nB(k(T))-f_a(k(T))\right],
        \label{eq:boltzmann}
\end{equation}
where $T_*=\tmin$ is our lower temperature. We have used that $\dd T=-3 T H(T) c_s^2(T)\dd t$ and we 
have rewritten it in terms of comoving momenta, defined in Eq.~\eqref{def_comoving}, to remove the
explicit Hubble term from Eq.~\eqref{defintrate}.
$H(T)=\sqrt{8\pi e(T)/(3m_\mathrm{Pl}^2)}$ is the Hubble rate --- $e(T)$ is the energy density
and $m_\mathrm{Pl}$ the Planck mass --- and $c_s^2(T)$ is the speed of sound squared. We follow the values tabulated in~\cite{Laine:2015kra}, which account for interactions, the electroweak transition and the quark mass thresholds.\footnote{%
\label{foot_sm}This only accounts for SM degrees of freedom. We thus neglect the axion's own contribution to the energy 
density of the universe. As we restrict ourselves to temperatures above the QCD transition, this  corresponds to
a percent-level uncertainty.
}

We choose our initial temperature $\tmax \ll \fPQ$ so that the effective
Lagrangian~\eqref{eq:laxion} is applicable. We also require that $\tmax$
is large enough that $\Gamma/H\gg 1$ for all $k$ there, so that axions are in thermal equilibrium.
Eq.~\eqref{eq:boltzmann} needs then to be integrated from the initial condition  $f_a(k(T=\tmax))=\nB(k(T=\tmax))$
to $\tmin$. If $\Gamma(k)/H(\tmin)\ll 1$  for all $k$ there and at all later temperatures, then the axion yield $Y_a=n_a/s$ or the comoving 
axion energy density\footnote{%
The normalization by the appropriate power of the entropy density was taken to factor out the effect 
of the expansion of the universe from the evolution of the values of $n_a$ and $e_a$.} 
$\eax=e_a/s^{4/3}$ will no longer evolve afterwards. Eq.~\eqref{eq:deltafirst} can then be rewritten as
\begin{equation}
    \Delta\neff=\frac{8}{7}\left(\frac{11}{4}\right)^{4/3}\frac{s^{4/3}(T_\mathrm{CMB})}{s^{4/3}(\tmin)}
    \frac{e_a(\tmin)}{e_{\gamma}(T_\mathrm{CMB})}.\label{eq:deltafirsts}
\end{equation}
We remark that the 
individual momentum modes are uncoupled in Eq.~\eqref{eq:boltzmann}, which can then be solved mode by mode. 
We can then integrate $f_a(k(\tmin))$ with respect to $k$ to obtain the energy density
and thence $\Delta \neff$ from Eq.~\eqref{eq:deltafirsts}. 
We will comment later on the choice of $\tmin$
with respect to the condition $\Gamma(k)/H(\tmin)\ll 1$ for all $k$.

In what follows we will not only quantify the theoretical uncertainty from the different schemes 
introduced in Sec.~\ref{sec:irdivergence},
but also that arising from adopting the commonly used \emph{momentum-averaged approximation}.
It follows by assuming  that, even during and after freeze out, axions are in kinetic equilibrium
with a multiplicative, temperature-dependent factor. While such an approximation 
is well justified in the classic case of WIMP freeze out \cite{Gondolo:1990dk}, where 
the processes responsible for chemical and kinetic equilibrium are different, it is not a 
priori clear if this is true in our case. 
Under that assumption, one sets $f_a(k(T))=f_a(T)\nB(k(T))$ and integrates the Boltzmann equation with respect to $k$.
One can then directly obtain an equation for $Y_a$
\begin{equation}
    \dv[Y_a(T)][\ln(T_*/T)]=\frac{\langle\Gamma\rangle (T)}{3 H(T) c_s^2(T)}\left[Y_\mathrm{eq}(T)-Y_a(T)\right],\label{eq:boltzmanny}
\end{equation}
and for $\eax$
\begin{equation}
    \dv[\eax(T)][\ln(T_*/T)]=\frac{\langle\Gamma k\rangle (T)\neq(T)}{3 H(T) c_s^2(T)\eeq(T)}\left[\tilde{e}_\mathrm{eq}(T)-\eax(T)\right],\label{eq:boltzmanne}
\end{equation}
where $Y_\mathrm{eq}(T)=\neq(T)/s(T)$ and $\tilde{e}_\mathrm{eq}(T)=\eeq(T)/s^{4/3}(T)$, with
$\neq(T)=\zeta(3)T^3/\pi^2$ and $\eeq(T)=\pi^2 T^4/30$. The definition of the momentum average 
$\langle\ldots\rangle$ has been given in Eq.~\eqref{defavg}.
The initial conditions are $Y_a(\tmax)=Y_\mathrm{eq}(\tmax)$ for \eqref{eq:boltzmanny} and  $\eax(\tmax)=\tilde{e}_\mathrm{eq}(\tmax)$ for \eqref{eq:boltzmanne}.
When solving Eq.~\eqref{eq:boltzmanne} one can again directly use its resulting value
of $e_a(\tmin)$ in Eq.~\eqref{eq:deltafirsts}. If, on the other hand, Eq.~\eqref{eq:boltzmanny} 
is solved, as in \cite{DEramo:2021psx,DEramo:2021lgb}, one needs to introduce
and determine the decoupling temperature $T_\mathrm{D}$.
It follows from 
the assumption
that the axion undergoes a freeze-out and decouples instantaneously from the thermal bath, 
at  $T=T_\mathrm{D}$. 
Under this approximation one finds (see e.g.~\cite{DEramo:2021lgb})
\begin{equation}
    \Delta\neff=\frac{4}{7}\left(\frac{11}{4}\right)^{4/3}\left(\frac{g^\mathrm{SM}_{*s}(T_\mathrm{CMB})}{g^\mathrm{SM}_{*s}(T_\mathrm{D})}\right)^{4/3},
    \label{eq:deltalast}
\end{equation}
where we have used the usual parametrization $s=2\pi^2 g_{*s} T^3/45$ of the entropy density, with 
$g_{*s}^\mathrm{SM}$ the effective number of entropy degrees of freedom in the SM. We employ the results from \cite{Laine:2015kra}; as in footnote~\ref{foot_sm}, we neglect the axion's
contribution to the entropy density.

Under this assumptions $Y_a$ stays equal to a limit value $Y_{\infty}$ it had when the axion decoupled from the thermal bath,
yielding
\begin{equation}
    Y_{\infty}=\frac{n_a(T_\mathrm{D})}{s(T_\mathrm{D})}=\frac{1}{s(T_\mathrm{D})}\int\frac{\dd^3 \vec{k}}{(2\pi)^3}\nB(k)=\frac{\zeta(3) T_\mathrm{D}^3}{\pi^2s(T_\mathrm{D})}
    =\frac{45\zeta(3)}{2\pi^4g_{*s}^\mathrm{SM}(T_\mathrm{D})}
    ,\label{eq:decdef}
\end{equation}
which  can be inverted numerically. The resulting value for $g_{*s}^\mathrm{SM}(T_\mathrm{D})$ can then be plugged in Eq.~\eqref{eq:deltalast}.

As we anticipated at the end of the previous section, 
we have chosen to avoid unphysical results by simply putting $\Gamma$ to $0$
whenever the computed value is negative. 
This applies to Equation \eqref{eq:boltzmann}, as well as to the computation of $\langle\Gamma\rangle$ and $\langle\Gamma k\rangle$ 
that enter Equation \eqref{eq:boltzmanny} and \eqref{eq:boltzmanne} respectively. As shown in Fig.~\ref{fig:controlfuncg3}, $\langle\Gamma\rangle$
in the strict LO scheme had been determined in \cite{Graf:2010tv} including the contribution of negative unphysical rates, so our treatment
here is slightly different.\footnote{%
\label{foot_neg}Including the negative contribution can be justified in the context of the momentum-averaged approximation, where 
it could even be compared against the result obtained by excluding nonphysical rates. However, as we aim 
at comparing the solutions of Eq.~\eqref{eq:boltzmann} with its momentum-averaged approximations
\eqref{eq:boltzmanny} and \eqref{eq:boltzmanne}, we consistently exclude
non-physical rates everywhere.
}

\begin{figure}[t]
    \centering
    \begin{subfigure}[b]{0.49\textwidth}
        \centering
        \includegraphics[width=\textwidth]{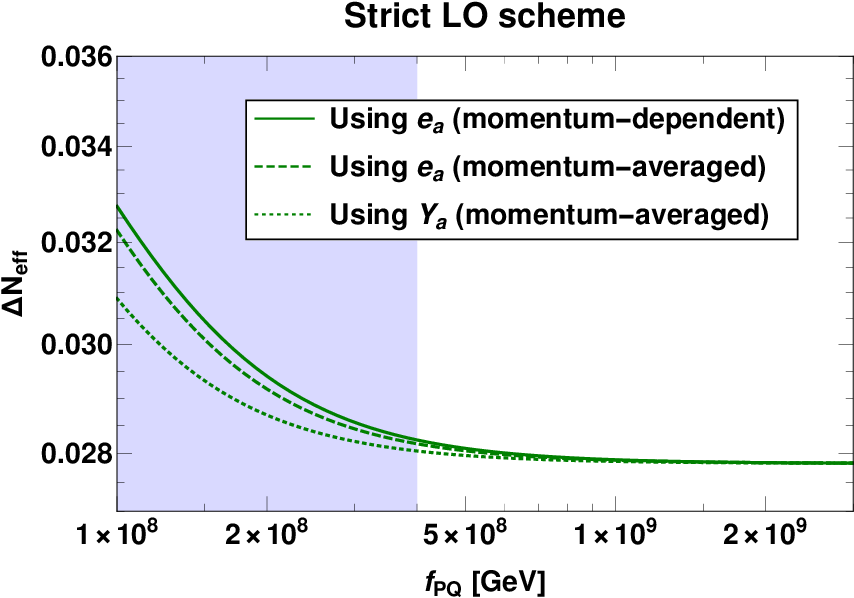}
        \caption{For the strict LO scheme}
    \end{subfigure}
    \hfill
    \begin{subfigure}[b]{0.49\textwidth}
        \centering
        \includegraphics[width=\textwidth]{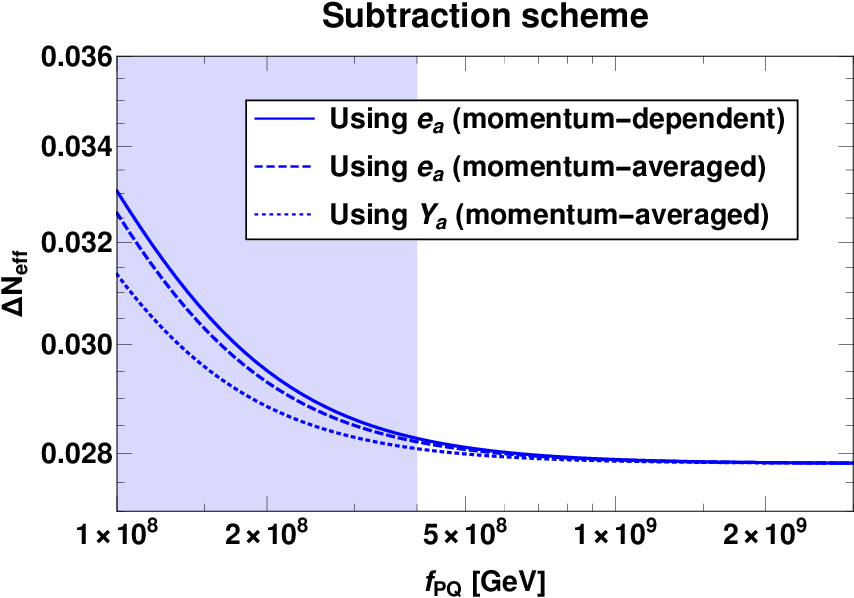}
        \caption{For the subtraction scheme}
    \end{subfigure}
    
    \bigskip
        \begin{subfigure}[b]{0.49\textwidth}
        \centering
        \includegraphics[width=\textwidth]{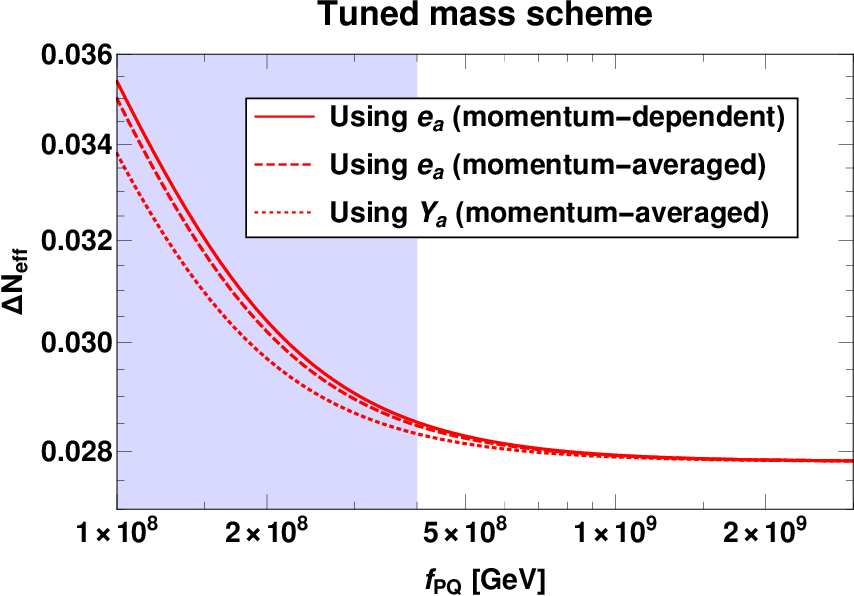}
        \caption{For the tuned mass scheme}
    \end{subfigure}
    
    \caption{Contribution to the effective number of neutrinos as a function of $\fPQ$. 
    Each plot uses a different scheme for the determination of the interaction rate. Within each scheme
    we compare the full solution of Eq.~\eqref{eq:boltzmann}, corresponding to the ``momentum-dependent'' label,
    with results coming from the solution of the momentum-averaged approximations for $\eax$, Eq.~\eqref{eq:boltzmanne},
    and for $Y_a$, Eq.~\eqref{eq:boltzmanny}. As explained in the main text, this last method requires the
    determination of  $T_\mathrm{D}$
    and relies on a very rapid freeze out. The shaded region corresponds to the values of $\fPQ$ excluded by the astrophysical limit in \cite{Carenza:2019pxu}. Please note that 
    the scale of the $y$ axis is the same on all plots.}
    \label{fig:deltaneff}
\end{figure}

We studied the dependence of $\Delta\neff$ on $\fPQ$
in the range above
$\fmin=10^8\;\mathrm{GeV}$.
This lower bound was chosen close to the $\fPQ \gtrsim 4\times 10^8\;\mathrm{GeV}$ bound from the observation of neutrinos emitted by SN1987A, as obtained in \cite{Carenza:2019pxu}.
See also \cite{Caputo:2024oqc} for a more recent and complete review of the current bounds on $\fPQ$, including the aforementioned \cite{Carenza:2019pxu}.
We note that all other bounds (\textit{e.g.} from neutron star cooling) are less constraining.
Figure \ref{fig:deltaneff} shows the final results for $\Delta\neff$.
As one can see, for values $\fPQ\gtrsim 10^9$ GeV the freeze out happens
well above the electroweak phase transition, where $\Delta\neff$ becomes essentially 
temperature-independent in a SM bath.

\begin{figure}[t]
    \centering
    \includegraphics[width=0.8\linewidth]{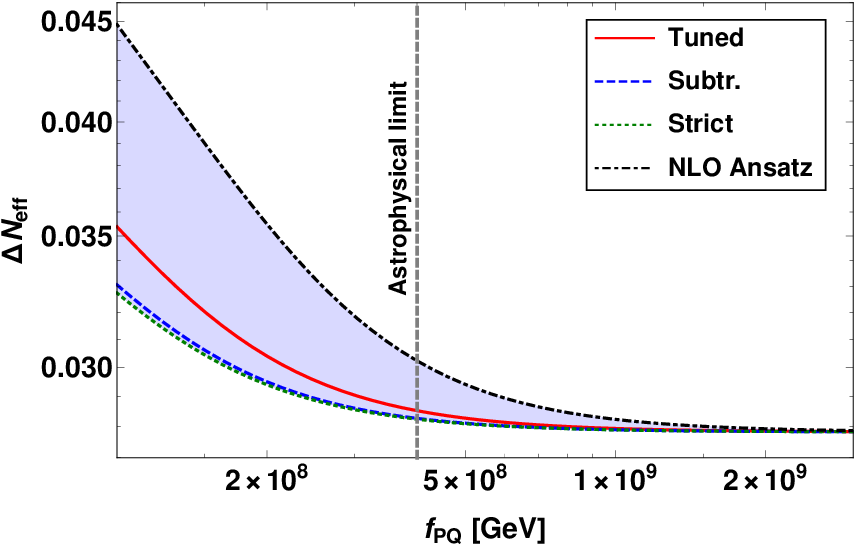}
    \caption{\kb{Contribution to the effective number of neutrinos as a function of $\fPQ$.
    All curves are obtained from the corresponding rates through the 
    full, momentum-dependent solution of Eq.~\eqref{eq:boltzmann}. 
    The shaded band corresponds to a conservative estimate of the contribution of unknown,
    potentially large NLO corrections to $\Gamma(k)_\text{KSVZ}$, which range
    from the strict LO to the NLO Ansatz given in Eq.~\eqref{largeNLO}.}}
    \label{fig:NLO}
\end{figure}

We further note that it was argued recently in~\cite{DEramo:2023nzt}
that, in non-axionic benchmark models for dark radiation, the absolute error on $\Delta\neff$ 
induced by using the momentum-averaged approximation instead of the momentum-dependent method may sometimes exceed the sensitivity of future surveys, a good example of which is CMB-S4 \cite{CMB-S4:2016ple,Abazajian:2019eic,CMB-S4:2022ght} which expects to be able to constrain $\Delta\neff < 0.06$ at $2\sigma$.
This is not reflected in our analysis. We find that the momentum-dependent and averaged determinations of $e_a$ within the same scheme
correspond to at most an absolute error of $5.0\times 10^{-4}$ for $\Delta \neff$.
On the other hand, the difference between momentum-dependent and $Y$-based determinations within the same scheme is at most $1.9\times 10^{-3}$. \kb{These values arise
from $\fPQ=10^8$ GeV. For the smallest allowed value  $\fPQ=4\times 10^8$ GeV
these numbers turn into $6.7\times 10^{-5}$ for the absolute error between the momentum-dependent and momentum-averaged determinations, and $1.9\times 10^{-4}$ for the absolute error between the momentum-dependent and $Y$-based determinations.}

\begin{figure}[t]
    \centering
    \begin{subfigure}[b]{0.48\textwidth}
        \centering
        \includegraphics[width=\textwidth]{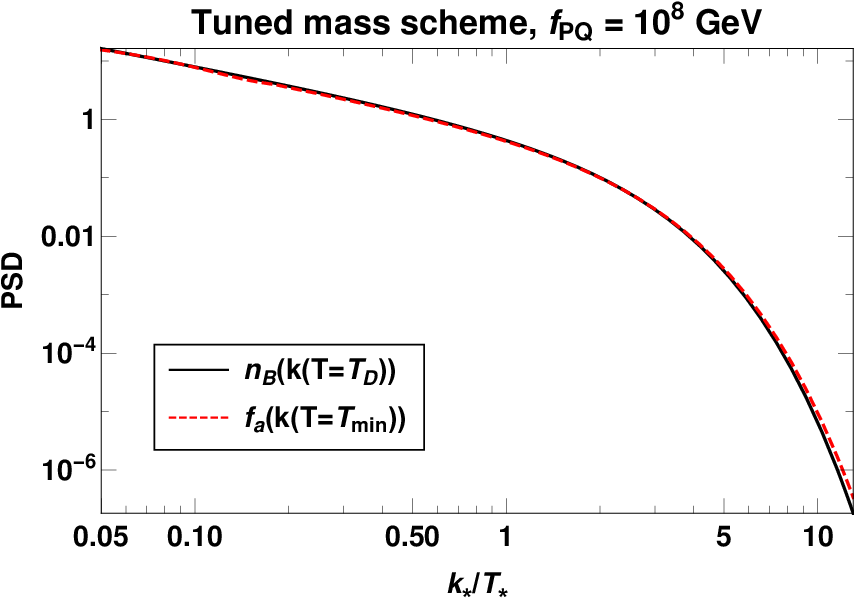}
        \caption{Comparison between the axion phase space distribution at $T=\tmin$ and the Bose--Einstein distribution at $T=T_\mathrm{D}$.}
        \label{fig:distortionf}
    \end{subfigure}
    \hfill
    \begin{subfigure}[b]{0.48\textwidth}
        \centering
        \includegraphics[width=\textwidth]{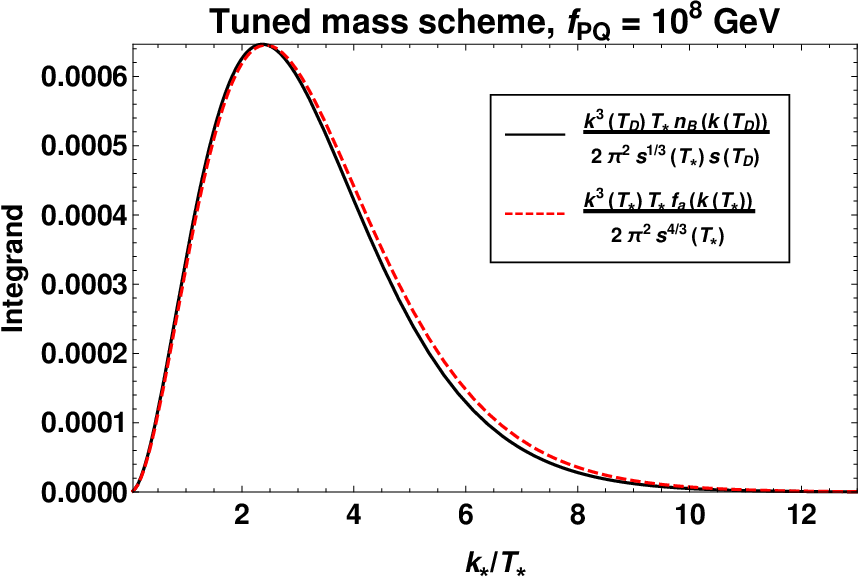}
        \caption{Integrand of $\eax$, compared between $T=\tmin$ and $T=T_\mathrm{D}$ where in the latter case $f_a$ has been replaced by $\nB$.}
        \label{fig:distortionintegrand}
    \end{subfigure}
    \caption{Effects of the spectral distortions on the distribution $f_a$ and on its third moment, which is proportional
    to the integrand for the comoving energy density $\eax$.
    }
      \label{fig:distortion}
\end{figure}

Figure~\ref{fig:deltaneff} also shows clearly that the choice of the computational scheme introduces a greater uncertainty than the error introduced by 
momentum averaging, as the difference between two schemes for the momentum-dependent method is at most $2.6\times 10^{-3}$ \kb{for $\fPQ=10^8$ GeV and $2.8\times 10^{-4}$ for  $\fPQ=4\times 10^8$ GeV. Fig.~\ref{fig:NLO} shows instead the effect of the potentially large NLO corrections on $\Delta\neff$. For the smallest allowed value $\fPQ=4\times 10^8$ GeV
the absolute spread between the strict LO and the NLO Ansatz is $2.0\times 10^{-3}$, corresponding to a 7\% effect on the former. As we push into lower, disallowed values of the axion
scale, we see that the band rapidly grows to a 40\% effect, to which 
the effect of delayed production at the QCD transition, discussed at the end of this 
section, would need to be added}.

\kb{Though the choice of computational scheme induces a larger, potentially much larger uncertainty than momentum averaging,} the momentum-dependent method is still better suited for precise computations as it accounts for 
\emph{spectral distortions}, i.e. momentum-dependent deviations from the equilibrium Bose--Einstein distribution 
that follow from different momenta decoupling at different times.
The momentum-averaged approximation, conversely, relies on the assumption that the axion phase space density can be written as $f_a(k,T)=\nB(k)f_a(T)$ and  does not account for spectral distortions.

In Figure~\ref{fig:distortion} we quantify the impact of spectral distortions. 
On the left we show  the axion phase space distribution $f_a$ at $\tmin$ and the (equilibrium) 
Bose--Einstein distribution at $T_\mathrm{D}$ for $\fPQ=10^8\;\mathrm{GeV}$.
As we can see the values of the two functions are very close, with the 
ratio $f_a(k(T=\tmin))/\nB(k(T=T_\mathrm{D}))$ being at most $\sim 1.9$.
This indicates the validity of the approximation 
that the axion phase space distribution gets ``frozen'' after decoupling as an equilibrium 
distribution at temperature $T_\mathrm{D}$, which in turn justifies the definition of the decoupling temperature.
The slight difference between the two curves at high $k_{*}/T_{*}$ are representative of the spectral distortions mentioned previously.
The difference between the two curves diminishes significantly as $\fPQ$ increases, in fact already at $\fPQ=10^9\;\mathrm{GeV}$ the two curves appear to perfectly overlap (in this case the ratio $f_a(k(T=\tmin))/\nB(k(T=T_\mathrm{D}))$ is at most $\sim 1.075$).

From Fig.~\ref{fig:distortionintegrand} we can understand that
spectral distortions are responsible from a feature clearly emerging from Fig.~\ref{fig:deltaneff}, i.e.
$\Delta \neff$ values computed accounting for momentum dependence  are always greater than those computed using
the momentum-averaged approximation. In Fig.~\ref{fig:distortionintegrand} we plot the integrand for the momentum integral in Eq.~\eqref{def_endensity}. One can then see how momentum modes at intermediate $k_*/T_*$, which
dominate this momentum integral, stay in equilibrium a little longer, thanks to their 
larger rate, and thus increase the final energy density and $\Delta\neff$ contribution.

\begin{figure}[t]
    \centering
    \begin{subfigure}[b]{0.48\textwidth}
        \centering
        \includegraphics[width=\textwidth]{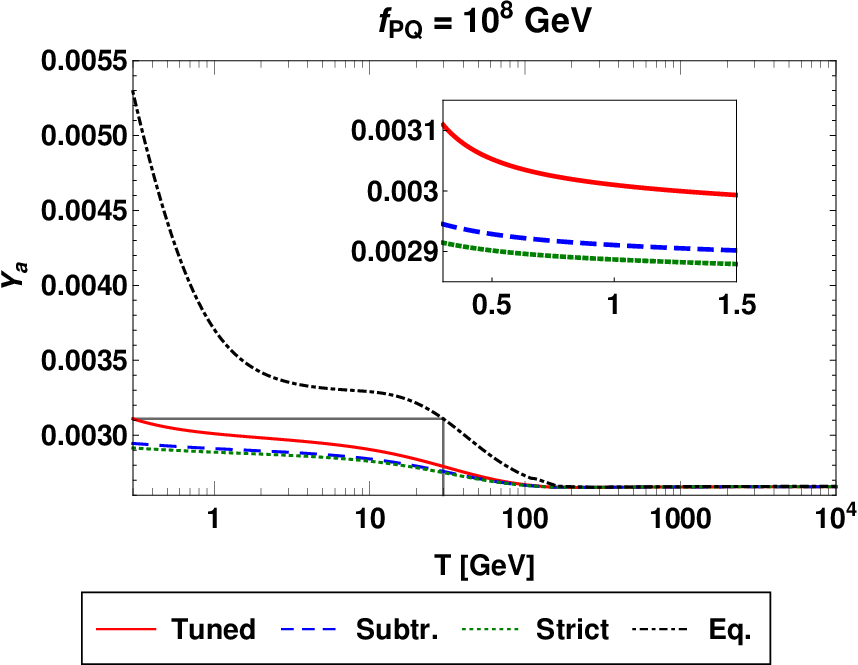}
        \caption{Axion yield}
    \end{subfigure}
    \hfill
    \begin{subfigure}[b]{0.48\textwidth}
        \centering
        \includegraphics[width=\textwidth]{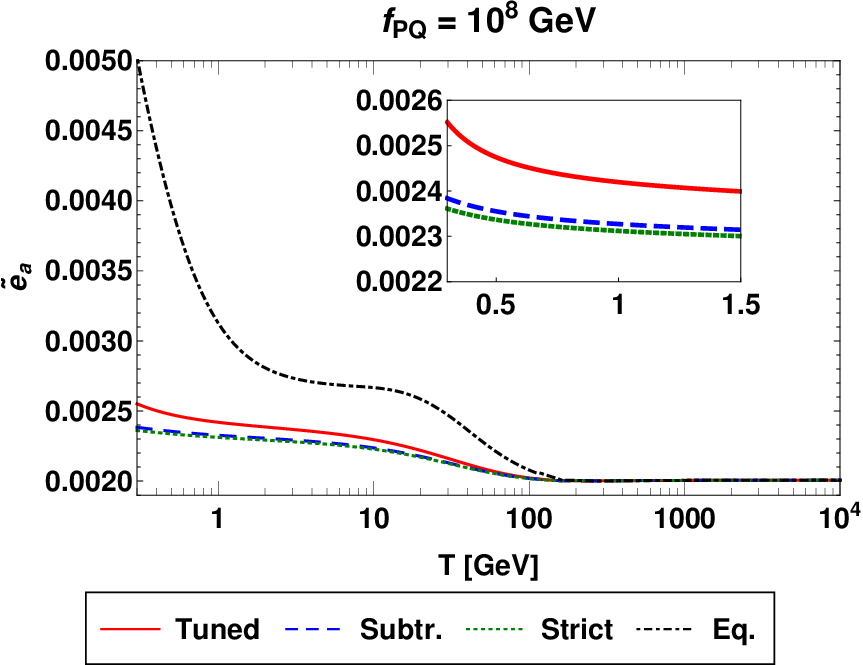}
        \caption{Axion energy density}
    \end{subfigure}  
    \caption{Yield and comoving energy density obtained from the momentum-dependent rates and $\fPQ=10^8\;\mathrm{GeV}$. The ``equilibrium'' value (``Eq.'') corresponds to
    $Y_\mathrm{eq}$ and $\tilde{e}_\mathrm{eq}$. The gray lines on the axion yield figure correspond to the graphical determination of $T_\mathrm{D}$, as defined by \eqref{eq:decdef}.}
    \label{fig:pheno}
\end{figure}

We remark that our determinations  of $\Delta \neff$ require that   $\eax(T)$ or $Y_a(T)$ are constant at $T=\tmin$ and below.
However, in some cases the slope of the curve continues to increase as $T\rightarrow\tmin$.
Figure~\ref{fig:pheno} shows $Y_a$ and $\eax$ for $\fPQ=10^8\;\mathrm{GeV}$.
The inset plots show the situation described earlier, especially for the tuned method.
As $T$ approaches $T_c\sim 155\;\mathrm{MeV}$, $g_3$ increases very rapidly which can counteract $\Gamma/H\sim \alpha_3^3 T m_\mathrm{Pl}/\fPQ^2$ 
dropping due to the overall $T^3$ driving the production rate.
This rise in $\Gamma/H$ signals the expected sensitivity 
to non-perturbative effects around the QCD transition, which might cause extra,
\kb{delayed} axion production,
as observed in~\cite{DEramo:2021lgb,DEramo:2021psx}. Hence, our current results should be 
considered as lower limits for $\Delta\neff$: a proper treatment requires a study of
axion production at and below the QCD phase transition. 
The methods presented in this article could be applied to this case as well, taking the Lagrangian of Chiral Perturbation Theory
\cite{Weinberg:1978kz,Gasser:1983yg,Gasser:1984gg} as  input. The obtained rate
would need to be merged with that in the perturbative regime, along the lines of~\cite{DEramo:2021lgb,DEramo:2021psx}.
We leave this to future work, \kb{together with the impact of the uncertainty
arising from the NLO Ansatz or from a proper NLO determination at QCD-transition temperatures.}

\section{Conclusions}
\label{sec_concl}
In this paper we have performed a careful analysis of the theory uncertainties associated with
thermal production of hot axions whose interactions with the early universe plasma
is dominated by their coupling to gluons. In Sec.~\ref{sub_naive}
we reviewed how a naive approach based on the standard Boltzmann equation with bare LO matrix elements
is insufficient. The well-known reason is that intermediate $t$- or $u$-channel gluons in these
matrix elements can have arbitrarily soft momenta; at these large wavelengths, they no longer
resolve the individual thermal particle constituents of the early-universe plasma.
This is signaled by a would-be infrared divergence, that is addressed by incorporating
collective plasma effects that arise at such long wavelengths. 

In Sec.~\ref{sub_resum} we showed how this was carried out in \cite{Graf:2010tv}
using Hard Thermal Loop resummation along the lines of~\cite{Braaten:1991dd}. This 
resulted in a \emph{strict LO} rate, given
in Eqs.~\eqref{strictLO} and \eqref{eq:prodratetotintstrict}. We then discuss
how this rate is valid for hard axions with $k\gtrsim T$ and how it extrapolates to 
non-physical negative rates for soft axions with $k\lesssim \mD\sim g_3T$. In App.~\ref{app_salvio}
we critically examine a more recent, gauge-dependent resummation scheme introduced in \cite{Salvio:2013iaa}, \jg{showing that, taken at face value, it gives rise 
to a divergent result. We argue that this issue also affects calculations
of gravitino \cite{Rychkov:2007uq} and axino \cite{Strumia:2010aa} production
which employ the same scheme.}
In Sec.~\ref{sub_schemes} we instead introduce two novel resummation schemes
inspired from right-handed neutrino and gravitational wave production on one 
hand and QCD thermalisation on the other. The former, called \emph{subtraction},
is defined in Eqs.~\eqref{subtrratefinal} and \eqref{eq:prodratetotintsubtr}. 
The latter is termed \emph{tuned mass} and it is given by Eqs.~\eqref{deftuned} 
and \eqref{eq:prodratetotinttuned}. It is constructed
to be gauge-invariant and positive-definite for all axion momenta and to 
reduce to the strict LO rate for $k\gtrsim T$ at small coupling. Thus, while still
an extrapolation in the soft momentum range, we argue it is a better suited one and we consider
it one of the main results of this paper. \kb{Given that the rate 
introduced in \cite{Salvio:2013iaa} is pathological, our tuned mass rate represents
the most refined determination of the axion rate in the KSVZ model.}

In Sec.~\ref{sec_prodrateres} we plot our numerical results for the 
strict LO, subtraction and tuned mass schemes. At large temperatures and hard axion momenta
they are close, with the spread between the tuned mass scheme and the other two larger than that between them. 
This is in agreement with our theoretical expectations and is shown by Figs.~\ref{fig:prodratetemp}, \jg{\ref{fig:diff}}
and \ref{fig:prodratemode}. These figures also show the unphysical negative turn of the strict LO and, 
to a lesser extent, subtraction schemes, which also causes the resulting rates for a given comoving 
momentum not to be a monotonic function of the temperature --- see Figs.~\ref{fig:prodratemode30}
and \ref{fig:prodratemode60}. 
The momentum-averaged rates plotted in Fig.~\ref{fig:controlfunc} show a similar behaviour:
the tuned mass rate is monotonic, increasing for decreasing temperature, while the tuned and strict LO
are not. While  all rates agree for very small couplings, as shown in Fig.~\ref{fig:controlfuncg3},
the spread between the more realistic tuned method and the other ones can be a factor of 5
close to the QCD transition. \kb{We further present an Ansatz for the axion rate 
at NLO in Eq.~\eqref{largeNLO}, built using NLO, $\mathcal{O}(g_3)$ corrections \cite{CaronHuot:2008ni} to a hot QCD 
quantity which shares the same operatorial form of the axion rate for soft gluon momenta.
This Ansatz is built to give  a conservatively large estimate of the size of NLO 
corrections, which are very large for this QCD quantity and related ones, as discussed 
in App.~\ref{app_QCD}. The momentum average of this NLO Ansatz rate can be a factor of 2 larger that the tuned mass one above the EW transition, growing to an almost tenfold
increase in the vicinity of the QCD transition.
}

To investigate the consequences of the rates determined from these three schemes,
in Sec.~\ref{sec_momdep} we use them to 
solve the momentum-dependent Boltzmann equation~\eqref{eq:boltzmann} and its 
momentum-averaged approximations for the yield $Y_a$ and comoving energy density $e_a/s^{4/3}$ 
in Eqs.~\eqref{eq:boltzmanny}-\eqref{eq:boltzmanne}. In Fig.~\ref{fig:deltaneff}
we show our results for $\Delta\neff$ as a function of the axion 
scale. For values of $\fPQ$ slightly smaller than astrophysically allowed ones 
and without considering any possible extra contribution from the QCD crossover and subsequent hadronic 
phase, we find that the absolute theory uncertainty from the scheme choice is of 
order $0.002$, while that coming from the popular momentum-averaged approximation 
is slightly smaller if one uses the integrated Boltzmann equation~\eqref{eq:boltzmanny} for $Y_a$. The error coming from the integrated Boltzmann equation~\eqref{eq:boltzmanne} for $e_a/s^{4/3}$ is instead a factor of five smaller, as one  is directly computing 
the energy density $e_a$ rather than having to infer it from 
the frozen-out $Y_a$ yield through a decoupling temperature.
\kb{Allowing for potentially large NLO, $\mathcal{O}(g_3)$ corrections through
our Ansatz, we find that the theory uncertainty from the scheme choice is $2.0\times 10^{-3}$ at
the minimum allowed value $\fPQ=4\times 10^8$ GeV, while it is only $2.8\times 10^{-4}$ if the NLO 
Ansatz is not included in the comparison. At lower, disallowed values
it grows to $0.012$. We thus consider $2.0\times 10^{-3}$ to be our most conservative 
estimate for the theory uncertainty to $\Delta\neff$ from astrophysically allowed KSVZ
axion dynamics above the QCD crossover. }

We consider this to be our \kb{other} main result, which we interpret as follows.
From a phenomenological point of view, the main axion dark radiation 
observable, $\Delta \neff$, is not particularly sensitive to 
differences in the rates. The reason is that axions freeze out
in the ultra-relativistic regime. There the equilibrium comoving number and energy
densities depend on the temperature only through changes in the appropriate 
power of $g_{*s}$ at the denominator, which never changes dramatically fast in a SM universe
in the region above the QCD crossover we explored here. Hence,
even rates differing by order 1 factors, as do our different schemes 
between the electroweak and QCD transition, can translate in the observed 
small differences in $\Delta\neff$, which in relative terms are just below 10\%
\kb{for the smallest allowed value of $\fPQ$}.
We would like to remark
that other approximations, such as neglecting the contribution 
of axions to the expansion rate --- see footnote~\ref{foot_sm} ---
should be smaller than this uncertainty.
We would also like to point out that the impact 
of the different schemes would be much larger if the axions 
were to never reach thermal equilibrium: it would thus be interesting 
to perform a similar analysis for axion-like particles with a freeze-in production
mechanism.

\begin{figure}[t]
    \centering
    \begin{subfigure}[b]{0.48\textwidth}
        \centering
        \includegraphics[width=\textwidth]{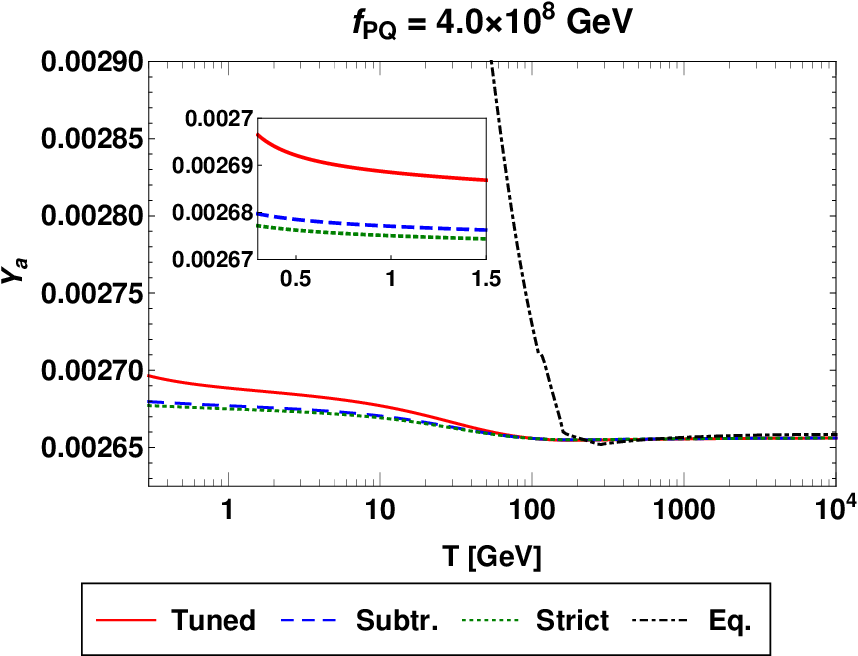}
        \caption{Axion yield}
    \end{subfigure}
    \hfill
    \begin{subfigure}[b]{0.48\textwidth}
        \centering
        \includegraphics[width=\textwidth]{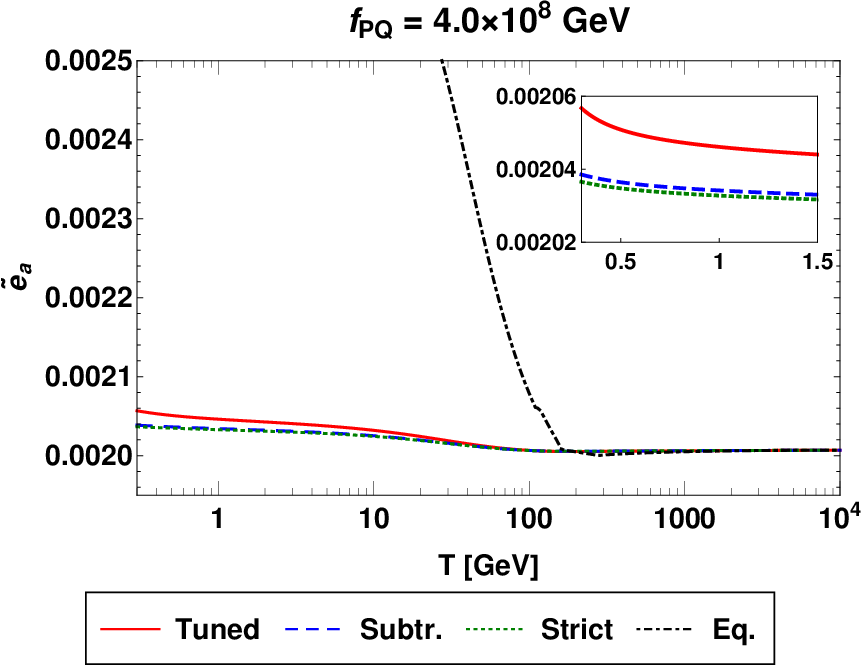}
        \caption{Axion energy density}
    \end{subfigure}  
    \caption{Yield and comoving energy density obtained from the momentum-dependent rates and $\fPQ=4.0\times 10^8\;\mathrm{GeV}$, close to the astrophysical bound of \cite{Carenza:2019pxu}. The ``equilibrium'' value (``Eq.'') is defined as in \ref{fig:pheno}.}
    \label{fig:pheno_astro}
\end{figure}

In Fig.~\ref{fig:pheno} we show the temperature evolution of the yield and comoving
energy density for the smallest value of $\fPQ$ we considered. It shows how
the resulting values, after a plateau in the 10 to 1 GeV range, start rising again 
when $T$ approaches the QCD crossover. Our results for $\Delta \neff$ miss the extra 
axion production happening below $T=300$ MeV and are thus to be considered as lower limits.
In Fig.~\ref{fig:pheno_astro} we plot the same quantities for the smallest
astrophysically allowed value of $\fPQ$, showing that the emergence of the effect 
of the QCD transition, though less marked, is still present. 
We leave the interpolation of our tuned mass scheme rate with a determination in the hadronic 
phase and the full quantification of this delayed, off-equilibrium production to future work.

\section*{Acknowledgements}
We are indebted to Mikko Laine for sharing with us the numerical results for the
running coupling and quark masses from \cite{Laine:2019uua}.
We would also like to thank him, 
Francesco D'Eramo, Fazlollah Hajkarim, Seokhoon Yun, Alberto Salvio, Alessandro Strumia, Wei Xue, Guy Moore, Greg Jackson and Giulio Alvise Dainelli
for useful conversations.
KB and JG acknowledge support by a PULSAR grant from the R\'egion Pays de la Loire.
JG is also funded
by the Agence Nationale de la Recherche under grant ANR-22-CE31-0018 (AUTOTHERM).

\appendix
\section{Phase-space integrations}
\label{app_phase}
In this Appendix we collect useful technical details on our implementation 
of phase-space integrations.
We remind the reader again that numerical results for the production rate, using the methods presented in this paper, are available on ZENODO \cite{ghiglieri_2024_10926565}.
Our starting point is the $2\leftrightarrow 2$ phase space, defined as
\begin{equation}
    \label{phase_space}
\int \! {\rm d}\Omega^{ }_{2\leftrightarrow 2}
 \equiv 
 \int \! \frac{{\rm d}^3\vec{p}_1^{ }
             \,{\rm d}^3\vec{p}_2^{ } 
             \,{\rm d}^3\vec{k}_1^{ }
              }
         {           (2 {p}^{ }_1)
                   \,(2 {p}^{ }_2)
                   \,(2 {k}^{ }_1)
           \, (2\pi)^9  }
  \, (2\pi)^4\, 
 \delta^{(4)}_{ }
                 \bigl(P^{ }_1 + P^{ }_2   - K^{ }_1 - K \bigr).
\end{equation}
In what follows we give more details on its implementation
in the different schemes.

We use the parametrization of \cite{Besak:2012qm}. In the 
$t$ channel it takes this form
\begin{equation}
    \label{tchannel}
     \int \! {\rm d}\Omega^{ }_{2\leftrightarrow 2}
 \; = \; 
 \frac{1}{(4\pi)^3_{ }k}
 \int^k_{-\infty} \! {\rm d}q^{ }_0 
 \int^{2k - q^{ }_0}_{|q^{}_0|} \! {\rm d}q 
 \int_{q_+^{ }}^\infty \! {\rm d}p^{ }_1 
 \int_{-\pi}^{\pi} \! \frac{{\rm d}\varphi}{2\pi}
 \;, \qquad
 q^{ }_{\pm} \; \equiv \; \frac{q^{ }_0 \pm q}{2}
 \;,
\end{equation}
where $t=q_0^2-q^2$, $K_1=P_1-Q$ and $P_2=K-Q$. This form 
emerges from using the energy-conserving $\delta$ functions to 
perform some angular integrations. This fixes the Mandelstam \kb{invariant} $s$ to
\begin{equation}
    \label{stchannel}
    s = 
 -\frac{t}{2q^2_{ }}\bigg[
   (2k - q^{ }_0)(2p^{ }_1 - q^{ }_0) + q^2_{ } 
  -\cos(\varphi)\, 
   \sqrt{(2 k - q^{ }_0 )^2_{ }-q^2}
   \sqrt{ (2 p^{ }_1 - q^{ }_0 )^2_{ }-q^2}\bigg]
 \;. 
\end{equation}
$u$ is easily obtained as $u=-s-t$.

In the $s$ channel the form above is inconvenient, as it 
would put cosines of the azimuthal angle at the denominator. 
It is more convenient to adopt this form
\begin{equation}
    \label{schannel}
     \int \! {\rm d}\Omega^{ }_{2\leftrightarrow 2}
 \; = \; 
 \frac{1}{(4\pi)^3_{ }k}
 \int_k^\infty \! {\rm d}q^{ }_0 
 \int_{|2k - q^{ }_0|}^{q^{}_0} \! {\rm d}q 
 \int_{q_-^{ }}^{q_+^{ }} \! {\rm d}p^{ }_2 
 \int_{-\pi}^{\pi} \! \frac{{\rm d}\varphi}{2\pi}
 \;, 
\end{equation}
where $s=q_0^2-q^2$, $P_1=Q-P_2$ and $K_1=Q-K$. The $t$ variable becomes
\begin{equation}
    \label{tschannel}
    t = 
 \frac{s}{2q^2_{ }}\bigg[
   (2k - q^{ }_0)(2p^{ }_2 - q^{ }_0) - q^2_{ } 
  + \cos(\varphi)\, 
   \sqrt{q^2 - (2 k - q^{ }_0 )^2_{ }}
   \sqrt{q^2 - (2 p^{ }_2 - q^{ }_0 )^2_{ }}\bigg]
 \;. 
\end{equation}

In the subtracted scheme we can easily take over the notation and results 
from \cite{Ghiglieri:2020mhm}. 
Let us define
\begin{equation}
{\rm I}\,{ }^t_{\sigma_1\sigma_2\tau_1}(a_1,a_2)=\int_{q_+^{ }}^\infty \! {\rm d}p^{ }_1 
 \int_{-\pi}^{\pi} \! \frac{{\rm d}\varphi}{2\pi}\left[a_1\frac{s^2+u^2}{t}+a_2t\right]\frac{n^{ }_{\sigma_1} (p^{ }_1)\,n^{ }_{\sigma_2}(p^{ }_2)\,[1 + n^{ }_{\tau_1}(k^{ }_1)]}{n_{\tau_1\sigma_1\sigma_2}(p^{ }_1 + p^{ }_2 - k^{ }_1)},
\label{eq:integrandt}
\end{equation}
with the expressions of $s$, $t$ and $u$ defined in \eqref{stchannel} and
\begin{equation}
{\rm I}\,{ }^s_{\sigma_1\sigma_2\tau_1}(b_1,b_2)=\int_{q_-^{ }}^{q_+^{ }} \! {\rm d}p^{ }_2\int_{-\pi}^{\pi} \! \frac{{\rm d}\varphi}{2\pi}\left[b_1\frac{t^2}{s}+b_2s\right]\frac{n^{ }_{\sigma_1} (p^{ }_1)\,n^{ }_{\sigma_2}(p^{ }_2)\,[1 + n^{ }_{\tau_1}(k^{ }_1)]}{n_{\tau_1\sigma_1\sigma_2}(p^{ }_1 + p^{ }_2 - k^{ }_1)},
\label{eq:integrands}
\end{equation}
with the expressions of $s$, $t$ and $u$ defined in \eqref{tschannel}.
$n_\sigma$ is defined as
\begin{equation}
n^{ }_\sigma(k) \; \equiv \; \frac{\sigma}{e^{k/T} - \sigma}
 \;, \quad
 \sigma = \pm\,,
\end{equation}
such that $n_{+}=\nB$ and $n_{-}=-\nF$.
The results from \cite{Ghiglieri:2020mhm} allow us to write\footnote{%
The decomposition used in \cite{Ghiglieri:2020mhm} for the graviton production rate integrand, in terms of $\tfrac{u^2+s^2}{t}$, $t$, $\tfrac{t^2}{s}$ and $s$ can be applied to the axion case because the dimension of the interaction term is the same, that is dimension $5$ suppressed by a mass scale. For other types of couplings a new basis would be needed.
}
for ${\rm I}\,{ }^t$
\begin{align}
{\rm I}\,{ }^t_{\sigma_1\sigma_2\tau_1}(a_1,a_2)&=\bigl[
  1 + n^{ }_{\tau_1\sigma_1}(q^{ }_0) 
    + n^{ }_{\sigma_2}(k - q^{ }_0) 
 \bigr] \, (q^2 - q_0^2)
 \nn  &\times
 \biggl\{
 \frac{a^{ }_1[q^2 - 3 (q^{ }_0 - 2k)^2]
 [12 L^{ }_3 + 6 q L^{ }_2 + q^2 L^{ }_1 ]}
 {6 q^4}
 - \biggl( a^{ }_2 + \frac{2 a^{ }_1}{3} \biggr) L^{ }_1
 \biggr\}\,,
\end{align}
where
\begin{align}
\label{l1}
L^{ }_1 
 & \equiv 
 T \Bigl[
   \ln\Bigl( 1 - \sigma^{ }_1 e^{-\qp^{ }/ T} \Bigr) 
 - 
   \ln\Bigl( 1 - \tau^{ }_1\, e^{\qm^{ }/ T} \Bigr) 
   \Bigr] 
 \;, \\
 \label{l2}
 L^{ }_2 
 & \equiv 
 T^2 \Bigl[
      {\rm Li}^{ }_2 
      \Bigl( \tau^{ }_1\, e^{\qm^{ }/ T} \Bigr) 
  - 
      {\rm Li}^{ }_2 
      \Bigl( \sigma^{ }_1 e^{ - \qp^{ }/ T} \Bigr) 
   \Bigr] 
 \;, \\ 
 \label{l3}
 L^{ }_3 
 & \equiv 
 T^3 \Bigl[
      {\rm Li}^{ }_3 
      \Bigl( \tau^{ }_1\, e^{\qm^{ }/ T} \Bigr) 
  - 
      {\rm Li}^{ }_3 
      \Bigl( \sigma^{ }_1 e^{ - \qp^{ }/ T} \Bigr) 
   \Bigr].
\end{align}
For ${\rm I}\,{ }^s$, we have
\begin{align}
{\rm I}\,{ }^s_{\sigma_1\sigma_2\tau_1}(b_1,b_2)&=\bigl[
  n^{ }_{\tau_1}(q^{ }_0 - k) - n^{ }_{\sigma_1\sigma_2}(q^{ }_0) 
 \bigr] \, (q^2 - q_0^2)
 \nn & \times 
 \biggl\{
 \frac{b^{ }_1  [q^2 - 3 (q^{ }_0 - 2k)^2]
 [12(L^{-}_3 - L^{+}_3) - 6 q(L^{-}_2 + L^{+}_2) 
 + q^2(L^{-}_1 - L^{+}_1) ]  }
 {12 q^4}
 \nn 
 &  \; - 
 \frac{b^{ }_1 (q^{ }_0 - 2k)
 [2(L^{-}_2 - L^{+}_2) - q(L^{-}_1 + L^{+}_1) ]  }
 {2 q^2}
 - 
 \biggl( \frac{b^{ }_1}{3} + b^{ }_2 \biggr)
 (L^{-}_1 - L^{+}_1 + q)
 \biggr\} \,,
\end{align}
where
\begin{align}
L^{\pm}_1 & \equiv  
 T \Bigl[
   \ln\Bigl( 1 - \sigma^{ }_1 e^{-\qmp^{ }/ T} \Bigr) 
 - 
   \ln\Bigl( 1 - \sigma^{ }_2\, e^{-\qpm^{ }/ T} \Bigr) 
   \Bigr] 
 \;, \\ 
 L^{\pm}_{2} & \equiv 
 T^2 \Bigl[
      {\rm Li}^{ }_2 
      \Bigl( \sigma^{ }_2\, e^{- \qpm^{ }/ T} \Bigr) 
  + 
      {\rm Li}^{ }_2 
      \Bigl( \sigma^{ }_1 e^{ - \qmp^{ }/ T} \Bigr) 
   \Bigr] 
 \;, \\ 
 L^{\pm}_{3} & \equiv 
 T^3 \Bigl[
      {\rm Li}^{ }_3 
      \Bigl( \sigma^{ }_2\, e^{ - \qpm^{ }/ T} \Bigr) 
  - 
      {\rm Li}^{ }_3 
      \Bigl( \sigma^{ }_1 e^{ - \qmp^{ }/ T} \Bigr) 
   \Bigr]\,.
\end{align}
Going back to Eqs.~\eqref{eq:prodrate_full} and \eqref{subtrratefinal}, we get
\begin{align}
\Gamma(k)^\text{subtr}_\text{KSVZ}=&\frac{g_3^6 (N_c^2-1)}{2^{13}\pi^7\fPQ^2 k^2}\left\{\int^k_{-\infty} \! {\rm d}q^{ }_0\int^{2k - q^{ }_0}_{|q^{}_0|} \! {\rm d}q
\bigg[N_c{\rm I}\,{ }^t_{+++}(-1,0)+2T_FN_f{\rm I}\,{ }^t_{-+-}(1,0)\right.\nn
&-\frac{4 \pi^2 T^3 k^2(q^2-q_0^2)}{q^4_{ }}\big(
    N_c
    +T_F N_f\big)\bigg]\nn
&+\left.\int_k^\infty \! {\rm d}q^{ }_0\int_{|2k - q^{ }_0|}^{q^{}_0} \! {\rm d}q\left[
N_c{\rm I}\,{ }^s_{+++}(-1,0)+T_FN_f{\rm I}\,{ }^s_{--+}(2,0)\right]\right\}\nn
&+\frac{g_3^4 (N_c^2-1) T\mD^2}{2^{10} \pi ^5 \fPQ^2}\ln\bigg(1+\frac{4k^2}{\mD^2}\bigg)\,,
\label{eq:prodratetotintsubtr}
\end{align}
where the  sign in front of ${\rm I}\,{ }^t_{-+-}$, which is the $q+g\to q+a$ contribution, comes from the fact that \eqref{eq:integrandt} contains $n_{-}(p_1)n_{+}(p_2)[1+n_{-}(k_1)=-\nF(p_1)\nB(p_2)[1-\nF(k_1)]$ which is the opposite sign compared to \eqref{eq:prodrate_full}. 
We have furthermore used the identity $\int \! {\rm d}\Omega^{ }_{2\leftrightarrow 2}(2t+s)\ldots=0$ to simplify terms where $\sigma_1=\sigma_2$.

Our implementation of the strict LO scheme, as explained in Sec.~\ref{sub_schemes}, is given 
by replacing $\ln(1+4k^2/\mD^2)$ with $\ln(4k^2/\mD^2)$, i.e.
\begin{align}
\Gamma(k)^\text{strict LO}_\text{KSVZ}=&\Gamma(k)^\text{subtr}_\text{KSVZ}-\frac{g_3^4 (N_c^2-1) T\mD^2}{2^{10} \pi ^5 \fPQ^2}\ln\bigg(1+\frac{4k^2}{\mD^2}\bigg)+\frac{g_3^4 (N_c^2-1) T\mD^2}{2^{10} \pi ^5 \fPQ^2}\ln\bigg(\frac{4k^2}{\mD^2}\bigg)\,.
\label{eq:prodratetotintstrict}
\end{align}

Regarding the tuned mass scheme, it is easy to see from \eqref{deftuned} that it amounts to making the following substitution in \eqref{eq:prodratetotintsubtr} 
\begin{equation}
{\rm I}\,{ }^t_{\sigma_1\sigma_2\tau_1}(a_1,a_2)\to \frac{q^4}{(q^2+\xi^2\mD^2)^2}{\rm I}\,{ }^t_{\sigma_1\sigma_2\tau_1}(a_1,-a_1/2)+{\rm I}\,{ }^t_{\sigma_1\sigma_2\tau_1}(0,a_2+a_1/2)\,,
\label{eq:substitutetuned}
\end{equation}
and undoing the subtraction of the bare term and the addition of the HTL-resummed last line, yielding
\begin{align}
\Gamma(k)^\text{tuned}_\text{KSVZ}=&\frac{g_3^6 (N_c^2-1)}{2^{13}\pi^7\fPQ^2 k^2}\left\{\int^k_{-\infty} \! {\rm d}q^{ }_0\int^{2k - q^{ }_0}_{|q^{}_0|} \! {\rm d}q
\bigg[\frac{N_cq^4}{(q^2+\xi^2\mD^2)^2}{\rm I}\,{ }^t_{+++}(-1,1/2)+N_c{\rm I}\,{ }^t_{+++}(0,-1/2)\right.\nn
&+2T_FN_f\frac{q^4}{(q^2+\xi^2\mD^2)^2}{\rm I}\,{ }^t_{-+-}(1,-1/2)
+2T_FN_f{\rm I}\,{ }^t_{-+-}(0,1/2)
\bigg]\nn
&+\left.\int_k^\infty \! {\rm d}q^{ }_0\int_{|2k - q^{ }_0|}^{q^{}_0} \! {\rm d}q\left[
N_c{\rm I}\,{ }^s_{+++}(-1,0)+T_FN_f{\rm I}\,{ }^s_{--+}(2,0)\right]\right\}.
\label{eq:prodratetotinttuned}
\end{align}
Readers familiar with \cite{York:2014wja} might wonder why we square $q^2/(q^2+\xi^2 \mD^2)$
in Eqs.~\eqref{deftuned} and \eqref{eq:substitutetuned}--\eqref{eq:prodratetotinttuned}. That is because the denominator structure of the bare diagram is a $1/t^2$
from the gluon propagators in the amplitude and conjugate amplitude, which gets softened to a $1/t$
by the derivatives in the axion-gluon coupling. As it is only the denominator that gets HTL-resummed,
we feel that our implementation is closer to that of  \cite{Arnold:2002zm,Arnold:2003zc}, even though a
non-squared factor of $q^2/(q^2+\xi^2 \mD^2)$ would also work, albeit with a different value of $\xi$. Attentive readers might have also noticed that our choice
causes the $N_c$-proportional part to become negative for $\mD/T\gg 1$, 
where $q^4/(q^2+\xi^2 \mD^2)^2\to 0$. A straightforward modification would be 
to replace 
\begin{equation}
    \frac{N_cq^4\,{\rm I}\,{ }^t_{+++}(-1,1/2)}{(q^2+\xi^2\mD^2)^2}+N_c\,{\rm I}\,{ }^t_{+++}(0,-1/2)\to \frac{N_cq^4\,{\rm I}\,{ }^t_{+++}(-1,1)}{(q^2+\xi^2\mD^2)^2}+N_c\,{\rm I}\,{ }^t_{+++}(0,-1)\,,
\end{equation}
in Eq.~\eqref{eq:prodratetotinttuned}. We have checked that this results in
a value of $F_3(\tmin)$ that is 5\% larger than  what we obtain from Eq.~\eqref{eq:prodratetotinttuned}. We employ the latter in our numerics to obtain
a marginally more conservative estimate of $\Delta\neff$.

The last remaining hurdle is to find the tuning coefficient $\xi$.
Let us now derive its numerical value.
The infrared-sensitive part of the $t$ channel integrand of 
$\Gamma(k)^\text{tuned}_\text{KSVZ}$ is obtained by 
removing the parts that are  proportional to $t$ only (because they vanish as $t\to 0$)
 and by expanding
 for $q_0,q\sim\mD\ll T$. It
reads 
\begin{equation}
\Gamma(k)^\text{tuned}_\text{KSVZ}\bigg\vert_{\text{soft}}=
\frac{g_3^6 (N_c^2-1)(N_c+T_FN_f)T^3}{2^{11}\pi^5\fPQ^2}\int^k_{-\infty} \! {\rm d}q^{ }_0\int^{2k - q^{ }_0}_{|q^{}_0|} \! {\rm d}q
\frac{q^2-q_0^2}{(q^2+\xi^2\mD^2)^2}+\ldots\,,
\label{eq:expandedtuned}
\end{equation}
where the dots stand for higher-order terms in the soft-$Q$ expansion.
Performing the integrals and expanding the obtained result for $k\gtrsim T$, we obtain
\begin{equation}
\Gamma(k)^\text{tuned}_\text{KSVZ}\bigg\vert_{\text{soft}}\approx
\frac{g_3^4(N_c^2-1)T\mD^2}{2^{10}\pi^5\fPQ^2}\bigg[\ln\left(\frac{k^2e^{2/3}}{\mD^2\xi^2}\right)+\mathcal{O}\left(\frac{\mD^2}{k^2}\right)\bigg],
\end{equation}
which we want to compare to the HTL contribution \eqref{HTLlimitsfinal} in the same limit, that is
\begin{equation}
\label{leadlog}
\Gamma(k\gtrsim T)_\text{KSVZ}^\text{HTL \cite{Ghiglieri:2016xye,Ghiglieri:2020mhm}}=\frac{g_3^4 (N_c^2-1) T\mD^2}{2^{10} \pi ^5 \fPQ^2}\bigg[\ln\bigg(\frac{4k^2}{\mD^2}\bigg)+\mathcal{O}\left(\frac{\mD^2}{k^2}\right)\bigg]\,.
\end{equation}
This gives $\xi^2=e^{2/3}/4$ or $\xi=e^{1/3}/2$ given that we chose $\xi$ to be positive.

As we mentioned in Sec.~\ref{sub_schemes} and \ref{sec_prodrateres}, the tuned
scheme differs from the strict LO and subtracted ones at relative order $g_3$. With 
the detailed results of this section we can determine this $\mathcal{O}(g_3)$ difference
analytically. Let us then inspect Eqs.~\eqref{eq:prodratetotintstrict} and 
\eqref{eq:prodratetotinttuned}. In particular, we need to concentrate on the region 
of the latter equation where $\vert q_0\vert\sim q\sim \mD$: as soon as these variables 
become larger than $\mD$, the mass-dependent denominators in Eq.~\eqref{eq:prodratetotinttuned}
can be expanded for $\mD/T\ll 1$, yielding deviations from the strict LO scheme 
that start at $\mathcal{O}(g_3^2)$. In the soft $\vert q_0\vert\sim q$ regime we can 
then expand Eq.~\eqref{eq:prodratetotinttuned} for  $\vert q_0\vert\sim q\sim \mD\ll T$
and consider the next order after Eq.~\eqref{eq:expandedtuned}. 

This next term can only come from scattering processes off gluons, as it requires a Bose--Einstein
enhancement from $\nB(p_1\sim gT)\approx T/p_1\sim 1/g$ to yield an order-$g_3$ term --- see
\cite{CaronHuot:2007gq,Caron-Huot:2008dyw,Fu:2021jhl} for similar calculations.
We then find
\begin{align}
 \nonumber   \Gamma(k)^\text{tuned}_\text{KSVZ}-\Gamma(k)^\text{strict LO}_\text{KSVZ}=&
    -\frac{g_3^6 (N_c^2-1) N_c T^2}{2^{12}\pi^7\fPQ^2 }\bigg\{\int^k_{-\infty} \! {\rm d}q^{ }_0\int^{2k - q^{ }_0}_{|q^{}_0|} \! {\rm d}q\bigg[\frac{1}{(q^2+\xi^2\mD^2)^2}-\frac{1}{q^4}\bigg]\\    
    \label{ogdiffstart}
&\times\frac{(q^2-q_0^2)(3 q q_0+(q^2-3q_0^2)\mathrm{arctanh}(q_0/q))}{q_0}+\mathcal{O}(g_3^2)\bigg\},
\end{align}
where the massless $q^4$ denominator on the first line comes from the strict LO scheme. 
The integration range can be split in the two domains $-q<q_0<q\land 0<q<k$ and 
$-q<q_0<2k-q\land q>k$. The latter range is negligible for $k\gtrsim T$, as the 
two denominators on the first line differ by $\mathcal{O}(g_3^2)$. We can then 
concentrate on the first range; we have exploited its symmetries to 
drop odd terms in $q_0$. Furthermore, at order $g_3$ we can extend the integration range to 
$-q<q_0<q\land 0<q$, finding that,
for $k\gtrsim T$
\begin{equation}
\label{ogdiff}
    \Gamma(k)^\text{tuned}_\text{KSVZ}-\Gamma(k)^\text{strict LO}_\text{KSVZ}=
    3(8+\pi^2)\frac{g_3^6(N_c^2-1)T^3}{2^{16}\pi^6\fPQ^2}\bigg[\frac{N_c\,\xi\mD}{T}+\mathcal{O}(g_3^2)\bigg].
\end{equation}

Let us now introduce our Ansatz for the NLO rate with the large 
$\mathcal{O}(g_3)$ corrections to the transverse momentum broadening 
coefficient $\hat{q}$ determined in \cite{CaronHuot:2008ni}.
This reference gives the soft contribution to $\hat{q}$  as
\begin{align}
    \frac{\hat{q}}{g_3^4 C_R T^3}=\frac{N_c+T_FN_f}{6\pi}\ln\frac{q^*}{\mD}+\frac{N_c}{6\pi}\frac{\mD}{T}
    \bigg[-\frac{3 q^*}{16 \mD}+3\frac{3\pi^2+10-4\ln 2}{16\pi}\bigg]\,,
    \label{nloqhat}
\end{align}
where $C_R=C_F$  for hard quark undergoing broadening and $C_R=C_A=N_c$ for a hard gluon
and $g_3T\ll q^*$ is the UV regulator on $q_\perp$ to isolate the soft contribution. 
The linear term in $q^*$ can be shown to cancel from an opposite contribution coming from the $q_\perp>q^*$ range.

We now assume that the $\mathcal{O}(g_3)$ correction to the axion rate is given by 
this $\mathcal{O}(g_3)$ correction to $\hat{q}$. As we shall explain in App.~\ref{app_QCD}, it can be considered  a sound estimate for the ``maximal'' size of NLO corrections,
given that $\hat{q}$ is the QCD observable featuring one of the largest known NLO corrections.
Upon rewriting Eq.~\eqref{leadlog} as
\begin{equation}
\label{axionLL}
\Gamma(k)_\text{KSVZ}^\text{LL}=\frac{g_3^6 (N_c^2-1) T^3}{2^{8} \pi ^4 \fPQ^2}\frac{N_c+T_FN_f}{6\pi}
\ln\frac{2k}{\mD}\,,
\end{equation}
where LL stands for leading logarithm,
and comparing it with Eq.~\eqref{nloqhat} we can find the overall
normalisation of $\hat{q}$ with respect to the axion rate.
We can then construct this Ansatz for the axion rate
under the hypothesis of $\hat{q}$-like large  $\mathcal{O}(g_3)$ corrections\footnote{%
\label{foot_linear}
\kb{For what concerns the linear term in $q^*$, the analysis we have carried out to obtain Eq.~\eqref{ogdiff} can be used to show that it cancels against an opposite term hidden
in the soft limit of the strict rate, i.e.
\begin{align}
 \nonumber  \Gamma(k)^{\text{strict LO }q^*}_\text{KSVZ}=&
    \frac{g_3^6 (N_c^2-1)  T^3}{2^{11}\pi^5\fPQ^2 }\bigg\{\int^{2k}_{q^*} \! {\rm d}q_\perp 
    \int^{\infty}_{-\infty} \! {\rm d}q^{ }_0
    \frac{q_\perp}{q} \frac{q_\perp^2}{q^4}\\    
    \label{linearterm}
&\times\bigg[N_c+T_F N_f-N_c\frac{(3 q q_0+(q^2-3q_0^2)\mathrm{arctanh}(q_0/q))}{2\pi^2 q_0 T}+\mathcal{O}(g_3^2)\bigg]\bigg\},\nn
=& \frac{g_3^6 (N_c^2-1)  T^3}{2^{8}\pi^4\fPQ^2 }\bigg[\frac{N_c+T_F N_f}{6\pi}\ln\frac{2k}{q^*}+\frac{N_c}{6\pi}\frac{3}{16T}(q^*-2k)\bigg]\,.
\end{align}
This indeed removes the linear dependence on the regulator, as expected. We have also
left for illustration the logarithmic one from the LO contribution.}}
\begin{equation}
    \label{largeNLO}
    \Gamma(k)_\text{KSVZ}^\text{NLO ansatz} \equiv \Gamma(k)^\text{strict LO}_\text{KSVZ}
    +\frac{g_3^6(N_c^2-1)T^3}{ 2^{8}\pi^4\fPQ^2} \frac{N_c}{2\pi}\frac{\mD}{T}
    \frac{3\pi^2+10-4\ln 2}{16\pi}.
\end{equation}
We could very well have used the tuned, rather than the strict, rate, as our starting point.
In that case we would have needed in principle to subtract its $\mathcal{O}(g_3)$ contribution, as given by Eq.~\eqref{ogdiff}. In practice this is irrelevant for 
$k\gtrsim T$, given that the latter 
is approximately 220 times smaller than the second therm on the r.h.s. of 
Eq.~\eqref{largeNLO}.

Finally, we remark that  the $q,q_0$ integrals  can be performed
numerically for all schemes to obtain the value of $\Gamma$ for a given $k,T$.
Note that for the subtracted and strict LO schemes the temperature dependence can be factored 
out of the integrals and it is therefore enough to construct an interpolator of the integral for the relevant values of $k/T$.
The only non-multiplicative dependence on the temperature enters through the 
logarithmic dependence on the coupling in the analytical term 
in Eqs.~\eqref{HTLlimitsfinal} and \eqref{eq:prodratetotintsubtr} and its strict LO 
limit~\eqref{eq:prodratetotintstrict} and is thus easily evaluated.
The same is not true for the tuned mass scheme where the integration must be done for all points of the grid (see Section \ref{sec_prodrateres}).
As the operation of computing the production rate over all $60000$ points of the grid can be very time-consuming, this task was performed in C using the GNU Scientific Library~\cite{Galassi:2019czg}.
The C implementation of the polylogarithm functions was taken from~\cite{voigt_2022_6951307}.

\kb{We conclude by listing} here the expressions for the Coulomb-gauge HTL-resummed gluon propagators, which have been employed in
Eqs.~\eqref{HTLfinal} and Eq.~\eqref{HTLlimits}. They read
\begin{eqnarray}
\label{htllong}
G^{00}_R(Q)&\equiv&G_{R}^L(Q)=\frac{i}{\displaystyle q^2+\mD^2\left(1-\frac{q^0}{2q}\ln\frac{q^0+q+i\epsilon}{q^0-q+i\epsilon}\right)},\\
\nonumber G^{ij}_R(Q)&\equiv&(\delta^{ij}-\hat q^i\hat q^j)G^T_R(Q)=
\left.\frac{i(\delta^{ij}-\hat q^i\hat q^j)}
     {\displaystyle q_0^2-q^2-m_\infty^2 \left(\frac{q_0^2}{q^2}
       -\left(\frac{q_0^2}{q^2}-1\right)\frac{q^0}{2q}
       \ln\frac{q^0{+}q}{q^0{-}q}\right)}\right\vert_{q^0=q^0+i\epsilon}.\\
&& 	\label{htltrans}
\end{eqnarray}
Here $m_\infty^2=\mD^2/2$ is the LO
gluon asymptotic mass.

\section{Gauge-dependent resummations in the literature}
\label{app_salvio}
Another purportedly positive-definite scheme was introduced in \cite{Salvio:2013iaa},
extending to the axion case the work of \cite{Rychkov:2007uq} for gravitinos 
and of \cite{Strumia:2010aa} for axinos. It
has been used in the analysis of \cite{DEramo:2021psx,DEramo:2021lgb}.
In a nutshell, the authors compute an analogue of Fig.~\ref{fig:axionself}
where both propagators are resummed. Adapting their Eqs.~(2.6) and (4.6) to our notation
and to the labeling of Fig.~\ref{fig:axionself} we find
\begin{equation}
\label{startsalvio}
    \Gamma(k)^\text{\cite{Salvio:2013iaa}}=\frac{(N_c^2-1)g_3^4}{2^8\pi^4k\fPQ^2}
    \int \frac{\dd^4Q}{(2\pi)^4}\epsilon^{\mu\nu\alpha\beta}\epsilon^{\mu'\nu'\alpha'\beta'}
    Q_{\alpha}Q_{\alpha'}(K-Q)_\beta(K-Q)_{\beta'} \frac{{}^*G^{<}_{\mu\mu'}(Q)\,{}^*G^<_{\nu\nu'}(K-Q)}{\nB(k)},
\end{equation}
where ${}^*G^{<}$ denotes the backward Wightman propagator, given by
${}^*G^<(Q)=\nB(q^0)\,({}^*G_R(Q)-{}^*G_A(Q))$.
${}^*G_R(Q)$ is  the resummed retarded propagator. When the
momenta $Q$ or $K{-}Q$ are space-like,  
the full one-loop Feynman-gauge retarded self-energy
is resummed. Conversely time- and light-like
momenta resum the self-energy in the HTL approximation, thus giving rise to zero-width
plasmons with momentum-dependent thermal masses.
This procedure is claimed to reduce to the strict LO 
at small $g_3$. 
It is further claimed that the gauge 
dependence arising from using the Feynman-gauge self-energy only
enters at relative order $g_3^2$.

We do not agree with these claims and we now carefully scrutinize them. Let us start by 
following \cite{Salvio:2013iaa} to turn Eq.~\eqref{startsalvio} into 
\begin{align}
    \Gamma(k)^\text{\cite{Salvio:2013iaa}}=&
   \frac{(N_c^2-1)g_3^4}{2^{13}\pi^7k^2\fPQ^2}
    \int_{-\infty}^\infty \mathrm{d}q^0\int_0^\infty \mathrm{d}q\int_{\vert q-k\vert}^{q+k}\mathrm{d}p 
   \, p q\, \frac{\nB(q^0)\nB(k-q^0)}{\nB(k)}\nonumber \\
   &\bigg\{\big(\rho^{FG}_L(Q)\rho^{FG}_T(P)+\rho^{FG}_T(Q)\rho^{FG}_L(P)\big)
   [(q+p)^2-k^2][k^2-(q-p)^2]\nonumber \\
   &\hspace{-1.5cm}+\rho^{FG}_T(Q)\rho^{FG}_T(P)\bigg[\left(\frac{q_0^2}{q^2}+\frac{p_0^2}{p^2}\right)
   \big((q^2+p^2-k^2)^2+4q^2p^2\big)+8q^0p^0(q^2+p^2-k^2)\bigg]
   \bigg\},\label{startsalvio2} 
\end{align}
where $p^0=k-q^0$ and $\rho^{FG}(Q)\equiv{}^*G_R(Q)-{}^*G_A(Q)$ is the Feynman-gauge
spectral density. 
In \cite{Salvio:2013iaa} these Feynman-gauge structures take this specific form
\begin{align}
\label{defrett}
    {}^*G^{T}_R(Q)&=\frac{i}{Q^2-\theta(-Q^2)\Pi^\text{vac}_R(Q)-\Pi_R^{T}(Q)},\quad\text{with}\quad 
    \Pi^T_R(Q)=1/2(\delta^{ij}-\hat{q}^i\hat{q}^j)\Pi^{ij}_R(Q)\,,\\
    \label{defrettr}
    {}^*G^{L}_R(Q)&=\frac{i Q^2}{q^2(Q^2-\theta(-Q^2)\Pi^\text{vac}_R(Q)-\Pi_R^{L}(Q))},
    \quad\text{with}\quad \Pi^L_R(Q)=-\frac{Q^2}{q^2}\Pi^{00}_R\,.
\end{align}
Here 
\begin{equation}
    \Pi^\text{vac}_R(Q)=-\bigg(\frac{5}{3} C_A-\frac43 T_FN_f\bigg) \frac{g^2 Q^2}{(4\pi)^2}\ln\frac{q^2-(q^0+i\epsilon)^2}{\bar\mu^2},
\end{equation}
is the vacuum component of the Feynman-gauge retarded self-energy. As per the prescription
of \cite{Salvio:2013iaa}, it is only used in the space-like regime, together with the full
Feynman-gauge thermal self-energies $\Pi^{L,T}_R(Q)$. Above the light cone
the HTL self-energies are used instead. In this appendix we will 
assume that any $\Pi$ self-energy is to be understood in Feynman gauge
unless otherwise specified.

Eq.~\eqref{startsalvio2} is then naturally divided into four components, 
which we name pole-pole, cut-cut, pole-cut and cut-pole. Pole-pole is the contribution
arising from $Q^2>0$ and $P^2=(K-Q)^2>0$, when both gluons are described
by their \emph{plasmon poles} encoded in the HTL spectral functions. Conversely,
the cut-cut emerges when $Q^2<0$ and $P^2=(K-Q)^2<0$, when both gluons are in 
the \emph{Landau cut}, i.e. the space-like branch cut of the thermal self-energies. Finally,
the pole-cut and cut-pole  are the two mixed contributions.

\begin{figure}[t]
    \centering
    \includegraphics[width=0.45\linewidth]{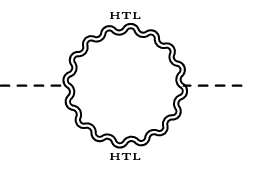}
     \includegraphics[width=0.35\linewidth]{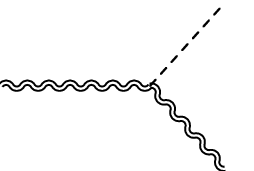}
    \caption{Left: the axion self-energy diagram corresponding
    to the pole-pole contribution. Its one cut corresponds
    to the square of the process shown on the right. See Sec.~\ref{sub_resum}
    for the relation between the retarded axion self-energy and the
    production rate.}
    \label{fig:polepole}
\end{figure}

We now consider the region where $k\gtrsim T$ at small $g_3$ and prove that 
Eq.~\eqref{startsalvio2} \emph{does not} reduce to the strict LO result. In this
region the pole-pole and cut-cut contributions are subleading. The former case
is depicted in Fig.~\ref{fig:polepole}. There the integral
in Eq.~\eqref{startsalvio2} is dominated by the region where $q\sim p\sim T$. Thus the longitudinal
pole contribution vanishes --- there are no hard longitudinal plasmons --- and 
the transverse one approaches a constant asymptotic mass $\mD^2/2\sim g_3^2 T^2$ \cite{Weldon:1982aq}. If the masses are exactly equal for the $Q$ and $K-Q$
zero-width transverse plasmons, no axion emission can happen for energy-momentum conservation.
Allowing for a small, momentum-dependent deviation between the two plasmon masses
gives rise to a suppressed contribution: the factor multiplying $\rho^{FG}_T(Q)\rho^{FG}_T(P)$
is of order $g_3^4$ or smaller. Combining this with
the multiplicative $g_3^4$ in the prefactor of Eq.~\eqref{startsalvio2}
we then have that
the pole-pole contribution is suppressed by
at least a factor of $g_3^2$ with respect to the strict leading order, which 
is $\sim g_3^6$.

\begin{figure}[t]
    \centering
    \includegraphics[width=0.45\linewidth]{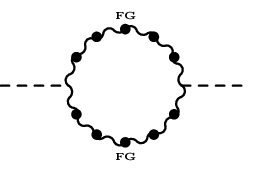}
     \includegraphics[width=0.35\linewidth]{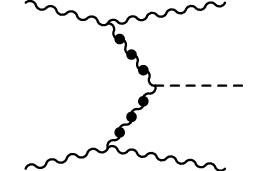}
    \caption{Left: the axion self-energy diagram corresponding
    to the cut-cut contribution. The black blobs denote the resummed 
    self-energies in Feynman Gauge (FG). Its  cut corresponds
    to the square of processes such as the one 
    shown on the right. Cuts corresponding to diagrams with external quarks
    are not shown.}
    \label{fig:cutcut}
\end{figure}

Similarly, the cut-cut contribution is also subleading, as it corresponds to higher-order 
$2\leftrightarrow 3$ processes, as shown in Fig.~\ref{fig:cutcut}. 
If both $P\sim T$ and $Q\sim T$, these processes are suppressed
by a factor of $g_3^2$ compared to leading order, since $\rho(Q\sim T)\sim
g^2/T^2$ in the space-like regime. For these processes to be 
enhanced to leading order we then need to look for regions where
either $Q$ or $P$ are soft. Let us look at the former case for illustration:
there one has $\rho(Q)\sim1/(g^2T^2)$ and $\rho^T(P)\sim g/T^2$. This latter
scaling can be understood as originating from $P=K+\mathcal{O}(gT)$, $P^2\sim gT^2$,
coupled with $\mathrm{Im}\Pi^T(P)\sim g^3T^2$ in this regime, as follows from 
Eq.~\eqref{transim}. Plugging these results into Eq.~\eqref{startsalvio2}
then yields that this soft-gluon contribution is also suppressed by a factor 
of $g_3^2$. Our numerical evaluation confirms this.\footnote{
A more careful power-counting analysis shows that the cut-cut contribution
should scale like $g_3^8\ln^2(1/g_3)$  in this non-abelian case and like 
$g_3^8\ln(1/g_3)$ in the abelian one, with the different power of the logarithm
originating 
from the extra logarithmic sensitivity to a collinear $P$ in the gluonic 
part of Eq.~\eqref{transim} compared with its quark counterpart.
Our numerical evaluation confirms this scaling.
}

\begin{figure}[t]
    \centering
    \includegraphics[width=0.45\linewidth]{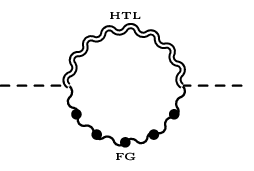}
     \includegraphics[width=0.35\linewidth]{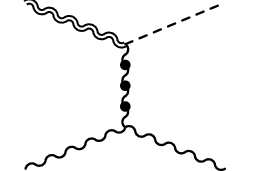}
    \caption{Left: the axion self-energy diagram corresponding
    to the pole-cut contribution. Its  cut corresponds
    to the square of processes such as the one 
    shown on the right. Cuts corresponding to external quarks are again not shown.}
    \label{fig:cutpole}
\end{figure}
So we are then left with the pole-cut and cut-pole contributions, as shown 
in Fig.~\ref{fig:cutpole}. These  correspond to 
the $2\leftrightarrow 2$ processes  we have treated in 
Sec.~\ref{sec:irdivergence}, with two notable differences. First,
the intermediate gluon resums a gauge-dependent self-energy, while one 
of the external gluons features the HTL, rather than bare, dispersion relation. Secondly, the 
crossing of the diagram shown on the right in Fig.~\ref{fig:cutpole}
into its $s$-channel counterpart
is not possible, as it would require a time-like $Q^2$. These two differences
are the basis of our claim that this gauge-dependent resummation does
not reproduce the strict LO result for $g_3\ll 1$.

To see this in detail, let us consider the sum of the pole-cut and cut-pole contributions, which 
is given by twice of either of the two.
 At small $g_3$ and for $k\gtrsim T$
we can again neglect the longitudinal plasmon. 
Hence
\begin{align}
    \Gamma(k)^\text{\cite{Salvio:2013iaa}}_\text{pc+cp}=&
   \frac{(N_c^2-1)g_3^4}{2^{12}\pi^7k^2\fPQ^2}
    \int_{-\infty}^\infty \mathrm{d}q^0\int_{\vert q^0\vert}^\infty \mathrm{d}q\int_{\vert q-k\vert}^{q+k}\mathrm{d}p 
   \, p q\,\big[1+\nB(q^0)+\nB(k-q^0)\big]\rho^{FG\text{ pole}}_T(P)\nonumber \\
  &\times \bigg\{\rho^{FG}_L(Q)
   [(q+p)^2-k^2][k^2-(q-p)^2]\nonumber \\
   &\hspace{-1.5cm}+\rho^{FG}_T(Q)\bigg[\left(\frac{q_0^2}{q^2}+\frac{p_0^2}{p^2}\right)
   \big((q^2+p^2-k^2)^2+4q^2p^2\big)+8q^0p^0(q^2+p^2-k^2)\bigg]\bigg\},\label{startsalviosanity} 
\end{align}
where we have taken $Q$ to be the cut momentum  and $P$ the pole one.
$\rho^{FG\text{ pole}}_T$ is the pole part of the transverse spectral function. It can again be approximated by 
its bare one, i.e.
\begin{equation}
    \rho^{FG\text{ pole}}_T(P\gg g_3T)\approx 2\pi\epsilon(k-q^0)\delta((k-q^0)^2-p^2)
    =\frac{\pi}{p}\epsilon(k-q^0)
    \big[\delta(p-\vert k-q^0\vert)+\delta(p+\vert k-q^0\vert)\big].
\end{equation}
Exploiting the fact that $Q^2<0$, we find that
only the $k-q^0>0$ region can contribute, yielding
\begin{align}
    \Gamma(k)^\text{\cite{Salvio:2013iaa}}_\text{pc+cp}=&
   \frac{(N_c^2-1)g_3^4}{2^{12}\pi^6k^2\fPQ^2}
    \int_{-\infty}^k \mathrm{d}q^0\int_{\vert q_0\vert}^{2k-q^0} \mathrm{d}q
   \,  q\, \big[1+\nB(q^0)+\nB(k-q^0)\big] (q^2-q_0^2)\nonumber \\
   &\times \bigg\{\rho^{FG}_L(Q)
   [(2k-q_0)^2-q^2]+\rho^{FG}_T(Q)\left(1-\frac{q_0^2}{q^2}\right)[(2k-q_0)^2+q^2]\bigg\}.\label{startsalviosanity2} 
\end{align}
In order to check that this equation misses the $s$-channel contribution,
as hinted by the lack of the corresponding crossing,
we proceed to take the spectral functions in the bare, one-loop limit, 
i.e. $\rho^{FG}_T(Q)\approx-2\mathrm{Im}\,\Pi_R^T(Q)/Q^4$ and similarly 
for $\rho^{FG}_L(Q)$. In this way we can use the known form of the thermal polarisation tensor
--- see e.g.~\cite{Kapusta:2006pm,Rychkov:2007uq} for the unintegrated Feynman-gauge
expressions and \cite{DEramo:2012uzl,Laine:2013vma} for the analytical momentum integration
yielding polylogarithms --- to find, for space-like $Q$
\begin{align}
\label{longsim}
    \mathrm{Im}\,\Pi_R^L(Q)=&\frac{g_3^2}{8\pi q^3}(q^2-q_0^2)\big[N_c(-q^2L_{1b}+4qL_{2b}+8 L_{3b})
    -8T_FN_f(qL_{2f}+2 L_{3f})
    \big],\\
       \mathrm{Im}\,\Pi_R^T(Q)=&\frac{g_3^2}{4\pi q^3}(q^2-q_0^2)\big[-N_c(q^2L_{1b}+qL_{2b}+2 L_{3b})
    +T_FN_f(q^2L_{1f}+2qL_{2f}+4 L_{3f}) \big],
    \label{transim}
\end{align}
where $L_{nb}\equiv L_{n}$ with $\tau_1=\sigma_1=1$ and $L_{nf}\equiv L_{n}$ with $\tau_1=\sigma_1=-1$, with $L_{n}$ given in Eqs.~\eqref{l1}--\eqref{l3}.
Upon plugging these expressions in Eq.~\eqref{startsalviosanity2} we
obtain
\begin{align}
    \Gamma(k)^\text{\cite{Salvio:2013iaa}}_\text{pc+cp}=&
   \frac{(N_c^2-1)g_3^6}{2^{13}\pi^7k^2\fPQ^2}
    \int_{-\infty}^k \mathrm{d}q^0\int_{\vert q_0\vert}^{2k-q^0} \mathrm{d}q
   \,   \big[1+\nB(q^0)+\nB(k-q^0)\big] (q^2-q_0^2)\nonumber \\
   &\times \bigg\{N_c \left[\frac{\left(3 (q_0-2 k)^2-q^2\right) \left(q^2L_{1b} +6
  q L_{2b} +12 L_{3b}\right)}{6 q^4}+\frac{5}{3} L_{1b}\right]\nonumber\\
   &\hspace{5mm}+2 T_F N_f \left[\frac{\left(q^2-3 (q_0-2 k)^2\right) \left( q^2L_{1f}+6
   qL_{2f} +12 L_{3f}\right)}{6 q^4}-\frac{2 }{3} L_{1f}\right]\bigg\}.\label{startsalviosanity3} 
\end{align}
Let us look back at Eq.~\eqref{eq:prodratetotintstrict}. The $t$-channel and $u$-channel
contribution thereto reads\footnote{This can be obtained by exploiting the 
definitions in Eqs.~\eqref{eq:integrandt}--\eqref{eq:integrands} and by noting that 
the proper separation of the $gg\to ga$ matrix element squared into the three
channels, as obtained  by explicit computation of the diagrams in Feynman gauge
as in \cite{Salvio:2013iaa}, reads
\begin{equation}
    |\mathcal{M}_{g+g\rightarrow g+a}|^2 = \frac{g_3^6(N_c^2-1)N_c}{32\pi^4\fPQ^2}\bigg[\underbrace{\frac{s u}{t}-t}_t+\underbrace{\frac{ st}{u}-u}_u
    +\underbrace{\frac{t u}{s}-s}_s\bigg].
     \label{matrixelementsrework}
\end{equation}
This is clearly equivalent to Eq.~\eqref{matrixelementsrework} and the $t$- and $u$-channel
contributions come from the terms thus labeled.
}
\begin{align}
\Gamma(k)^\text{strict LO t+u}_\text{KSVZ}=&\frac{g_3^6 (N_c^2-1)}{2^{13}\pi^7\fPQ^2 k^2}\int^k_{-\infty} \! {\rm d}q^{ }_0\int^{2k - q^{ }_0}_{|q^{}_0|} \! {\rm d}q
\bigg[-N_c{\rm I}\,{ }^t_{+++}(1,1)+2T_FN_f{\rm I}\,{ }^t_{-+-}(1,0)\nn
&-\frac{4 \pi^2 T^3 k^2(q^2-q_0^2)}{q^4_{ }}\big(
    N_c
    +T_F N_f\big)\bigg]
+\frac{g_3^4 (N_c^2-1) T\mD^2}{2^{10} \pi ^5 \fPQ^2}\ln\bigg(\frac{4k^2}{\mD^2}\bigg)\,,
\label{eq:prodratetotintsubtrt}
\end{align}
Neglecting for a moment the subtraction of the IR-divergent part and its replacement 
by the analytical result in Eq.~\eqref{HTLlimitsfinal}, we see that Eq.~\eqref{startsalviosanity3}
agrees with Eq.~\eqref{eq:prodratetotintsubtrt}, thus showing the lack of the $s$-channel
contribution in the small $g_3$ limit of the results of \cite{Salvio:2013iaa}.\footnote{The $s$-channel part of the matrix
elements and the corresponding diagrams appear in intermediate 
stages of the derivation of \cite{Salvio:2013iaa}. However, they are then 
subtracted, see Eq.~(2.8) and Table 1 there. If the pole-pole 
contribution were to include a one-loop width, their method would account for
them.}

What we have discussed is in our opinion not the main reason for which the method of 
\cite{Salvio:2013iaa} does not reproduce the strict LO result at small $g_3$. To explain 
why, we now need to turn to the IR sector of the cut-pole and pole-cut expression~\eqref{startsalviosanity}. For illustration purposes we keep using the
bare spectral function for the transverse pole, 
leading to Eq.~\eqref{startsalviosanity2}.
Let us now inspect the soft intermediate-gluon sector: for $q_0,q\sim g_3T\ll k$,
we find
\begin{align}
    \Gamma(k)^\text{\cite{Salvio:2013iaa}}_\text{pc+cp soft}=&
   \frac{(N_c^2-1)g_3^4}{2^{10}\pi^6\fPQ^2}
    \int_{\mathcal{O}(g_3T)} \mathrm{d}q^0 \mathrm{d}q
   \,  q\, \frac{T}{q_0} (q^2-q_0^2)
   \left[\rho^{FG}_L(Q)+\rho^{FG}_T(Q)\left(1-\frac{q_0^2}{q^2}\right)\right].\label{startsalviosanitysoft} 
\end{align}
Here $\int_{\mathcal{O}(g_3T)}$ means that this form of the integrand is valid for $q_0,q\sim g_3T$.
As we have explained in Sec.~\ref{sub_resum}, the strict LO result emerges from using gauge-invariant HTL-resummed spectral functions. Are the Feynman-gauge expressions above equivalent
to these in that limit at small $g_3$? Naively, one would expect that to be the case,
since HTLs are precisely the $\mathcal{O}(g_3^2T^2)$ 
gauge-invariant limit of thermal polarization tensors for soft
external momenta. Indeed, Eq.~\eqref{startsalviosanitysoft} would
agree with Eq.~\eqref{HTLfinal} if the soft-$Q$ Feynman-gauge
spectral functions were to agree with the HTL ones.

There is however a subtlety related to the transverse spectral function,
which in a generic space-like regime reads, following from Eq.~\eqref{defrettr}
\begin{equation}
    \label{spfret}
   \rho^{FG}_T(Q)=\frac{-2\mathrm{Im}\,\Pi_R^T(Q)}{(Q^2-\Pi^\text{vac}_R(Q)-\mathrm{Re}\,\Pi_R^T(Q))^2+(\mathrm{Im}\,\Pi_R^T(Q))^2}.
\end{equation}
The subtlety is that, though parametrically $\mathrm{Re}\,\Pi_R^T(Q\sim g_3T)=
\mathrm{Re}\,\Pi_R^{T\text{ HTL}}(Q)+\mathcal{O}(g_3^3T^2)$, there is a region where
the real part of the transverse HTL self-energy vanishes and the subleading, gauge
dependent term takes over. This happens when $q_0\to 0, q\lesssim g_3T$. 
In this region
\kb{\begin{equation}
\label{impartTsmallfreq}
  \mathrm{Im}\,\Pi_R^T(q_0\to 0, q\lesssim g_3T)  =
  -\frac{\pi  \mD^2 q^0}{4 q}-\frac{ g_3^2 N_c q^0 T}{4 \pi }+\mathcal{O}(g_3^2 q_0^3 T^2/q^2,g_3^2 q^0 q)\,,
\end{equation}
where the first term on the r.h.s. is the HTL term and the second 
is its first correction. Hence}
$\mathrm{Im}\,\Pi_R^T(Q)\propto q^0$, so that it vanishes at the denominator of 
Eq.~\eqref{spfret}, whereas at the numerator the power of $q^0$ is absorbed 
by the $T/q^0$  in Eq.~\eqref{startsalviosanitysoft}, which stems
from the classical-field tail of the Bose--Einstein distribution
$\nB(q^0)$. 
The real part reads instead --- see \cite{Kalashnikov:1980tk,Kajantie:1982xx} 
for the linear term in $q$
\begin{align}
    \Pi^\text{vac}_R(0,q)+\mathrm{Re}\,\Pi_R^T(0,q\ll T)=&-\frac{3 g_3^2 N_c q T}{16}
    +
    \frac{g_3^2 q^2}{(4\pi)^2} \left(\frac53 N_c-\frac43 T_F N_f\right) \ln\frac{(4\pi T)^2}{e^{2\gamma_E}\bar\mu^2}
    \nonumber \\
    &+
    \frac{g_3^2 q^2}{3(4\pi)^2}\bigg[\frac{28}{3}N_c
    +4T_FN_F\bigg(2\ln4-\frac{5}{3}\bigg )\bigg]+\ldots\,,
    \label{trouble}
\end{align}
where $\gamma_E$ is the Euler--Mascheroni constant. This expression shows 
clearly how in this region the HTL vanishes altogether: there is no $g_3^2T^2$
term. Furthermore, though we have kept only the first two terms of the $q/T$
expansion, we find that the expression~\eqref{trouble} describes well the full
numerical expression even at $q\approx 2 T$. 

We are then at the crux  of the problem: Eq.~\eqref{trouble} is negative both 
in the numerical form and in the approximate one.
When plugged in the spectral function in Eq.~\eqref{spfret} it creates,
together with the vanishing width at the denominator, a pole on the integration range.
Let us keep for illustration the leading, linear term in Eq.~\eqref{trouble}
and plug it in Eq.~\eqref{startsalviosanitysoft}: the divergent contribution is then
\kb{\begin{align}
   \label{startsalviosanityboomstart}
  \Gamma(k)^\text{\cite{Salvio:2013iaa} div}_\text{pc+cp}\approx &
   \frac{(N_c^2-1)g_3^4 T \mD^2}{2^{11}\pi^5\fPQ^2}
    \int_{\mathcal{O}(g_3^2 T)} \mathrm{d}q \int_{-q_0*}^{q_0*} \mathrm{d}q^0
   \,   \frac{ 1+\mathcal{O}(q/T)}{\left(q-\frac{3}{16} g_3^2 N_c T\right)^2+\pi^2\mD^4 \frac{q_0^2}{16q^4}}\\
    \label{startsalviosanityboommid}
   =& \frac{(N_c^2-1)g_3^4 T }{2^{8}\pi^6\fPQ^2}
    \int_{\mathcal{O}(g_3^2 T)}
   \frac{ \mathrm{d}q \,q^2}{q-\frac{3}{16} g_3^2 N_c T} 
   \arctan\frac{q_0^* \pi\mD^2/4}{q^2 \left(q-\frac{3}{16} g_3^2 N_c T\right)}\\
      \label{startsalviosanityboom}
       \approx & \frac{9(N_c^2-1)N_c^2 g_3^8 T^3 }{2^{17}\pi^5\fPQ^2}
    \int_{\mathcal{O}(g_3^2 T)} \frac{ \mathrm{d}q}{\left\vert q-\frac{3}{16} g_3^2 N_c T\right\vert}\to \infty\,,
\end{align} where for illustration we replaced only the HTL term  of Eq.~\eqref{impartTsmallfreq}
in Eq.~\eqref{spfret}. $g_3^4T\ll q_0*\ll g_3^2 T$ is a cutoff that singles out
the divergent region where these approximations are valid.}

\kb{Eqs.~\eqref{startsalviosanityboomstart}--\eqref{startsalviosanityboom}} then show how the gauge-dependent resummation of \cite{Salvio:2013iaa,Rychkov:2007uq}, even
when taken at small $g_3$, results in a divergent contribution. We have checked numerically
that this divergence persists when the full form of the HTL time-like spectral functions and 
of the Feynman-gauge resummed spectral function~\eqref{spfret} are used, both for $g_3\ll 1$
and $g_3\gtrsim 1$. It is not clear to us how this divergence was regulated in the numerical
results of \cite{Salvio:2013iaa}, but we believe it to be the main source
of the $\mathcal{O}(10)$ discrepancy between our results and theirs for $F_3$ discussed
in and around Fig.~\ref{fig:controlfunc}.  
We can speculate that, if \kb{the analytical expressions~\eqref{longsim}--\eqref{transim}
 for the imaginary parts of the self-energies are replaced by} interpolators, \kb{these}
 might generate a nonvanishing $\mathrm{Im}\,\Pi^T_R(0,q)$. \kb{This}
 would \kb{in turn} artificially regulate Eq.~\eqref{startsalviosanityboom} and inflate
the result obtained from this method when compared to the ones 
we present in the main text.\footnote{%
\label{foot_interpolator}
 Let us assume for argument's sake that this interpolating procedure introduces
a minimum value $\gamma^2_\text{num}>0$ for $(\mathrm{Im}\,\Pi^T_R(q^0\to 0,q))^2$.
Then the spectral-function denominator in Eq.~\eqref{startsalviosanityboomstart}
turns into $(q-3 g_3^2 N_c T/16)^2+\text{min}(\pi^2\mD^4 q_0^2/(16q^2),\gamma^2_\text{num})/q^2$. Let us assume that
the switchover between the proper, odd imaginary part and the artificial 
numerical width $\gamma_\text{num}$ happens for
 frequencies $\tilde{q}^0=\pm 4\gamma_\text{num} q/(\pi\mD^2)$ obeying $\tilde{q}_0^2<q_0^{*2}$. Then, 
 restricting ourselves to the frequency range $q_0^2<\tilde{q}_0^2$,
this numerically-regulated result differs from 
Eqs.~\eqref{startsalviosanityboomstart}--\eqref{startsalviosanityboom}.
It reads 
\kb{\begin{align}
    \Gamma(k)^\text{\cite{Salvio:2013iaa} div num}_\text{pc+cp}=&
   \frac{(N_c^2-1)g_3^4 T \mD^2}{2^{11}\pi^5\fPQ^2}
    \int_{\mathcal{O}(g_3^2 T)} \mathrm{d}q \int_{-\tilde{q}_0}^{\tilde{q}_0} \mathrm{d}q^0
   \,   \frac{ q^2}{\left(q^2-\frac{3}{16} g_3^2 N_c T q\right)^2+\gamma^2_\text{num}}\nonumber \\
   =& \frac{(N_c^2-1)g_3^4 T\gamma_\text{num} }{2^{8}\pi^6\fPQ^2}
    \int_{\mathcal{O}(g_3^2 T)}
\frac{ \mathrm{d}q \,q^3}{(q^2-\frac{3}{16} g_3^2 N_c Tq)^2+\gamma^2_\text{num}}
      \label{startsalviosanityboomnum}
       \approx 
       \frac{9(N_c^2{-}1)N_c^2 g_3^8 T^3 }{2^{16}\pi^5\fPQ^2}\bigg[1
       +\mathcal{O}\bigg(\frac{\gamma_\text{num}}{g_3^4T^2}\bigg)\bigg],
\end{align}}
where we performed the $q$ integration over the pole region, exploiting 
its Lorentzian-like shape.  This artificially regulated quantity is finite,
unlike Eq.~\eqref{startsalviosanityboom}. We remark that we have 
introduced a numerical width $\gamma_\text{num}$ only in the denominator
of Eq.~\eqref{spfret}. Were we to assume the same for the numerator, we would need to multiply the 
r.h.s. on the first line of Eq.~\eqref{startsalviosanityboomnum} by 
$\gamma_\text{num}/(\pi \mD^2 q^0/(4q))$, which would give a vanishing contribution if the $dq^0/q^0$
integration is treated as a principal value.

Les us further note that in Eq.~\eqref{startsalviosanityboomnum} we only
considered the contribution from $0<q_0^2<\tilde{q}_0^2$, which is 
where the divergence shows up in the absence of numerical regularization.
The range $q_0^{*2}>q_0^2>\tilde{q}_0^2$ depends then much more strongly on $\gamma_\text{num}$
--- logarithmically, in fact ---
than Eq.~\eqref{startsalviosanityboomnum}, i.e.
\kb{\begin{align}
    \Gamma(k)^\text{\cite{Salvio:2013iaa} div num remainder}_\text{pc+cp}=&
   \frac{(N_c^2-1)g_3^4 T \mD^2}{2^{11}\pi^5\fPQ^2}
    \int_{\mathcal{O}(g_3^2 T)} \mathrm{d}q  \int_{\tilde{q}_0}^{q_0^*} \mathrm{d}q^0
   \,   \frac{ 2}{\left(q-\frac{3}{16} g_3^2 N_c T\right)^2+\pi^2\mD^4 \frac{q_0^2}{16q^4}}\nonumber \\
   & \hspace{-3cm}=\frac{(N_c^2{-}1)g_3^4 T }{2^{8}\pi^6\fPQ^2}
    \int_{\mathcal{O}(g_3^2 T)}
   \frac{ \mathrm{d}q \,q^2}{q-\frac{3}{16} g_3^2 N_c T} 
  \Bigg[ \arctan\left(\frac{q_0^* \pi\mD^2/4}{q^2 \left(q-\frac{3}{16} g_3^2 N_c T\right)}\right)
  -\arctan\left(\frac{\gamma_\text{num}}{q \left(q-\frac{3}{16} g_3^2 N_c T\right)}\right)\Bigg]\nonumber \\
      \label{startsalviosanityboomnumlog}
       \approx & \frac{9(N_c^2{-}1)N_c^2g_3^8 T^3 }{2^{16}\pi^5\fPQ^2} \ln\frac{4q_0^*\pi\mD^2}{ 3 g_3^2 N_c T \gamma_\text{num}}.
\end{align}}
It is this logarithmic dependence on the regulator that would be responsible, under this hypothesis,
for an unphysical upward shift of the rate.
}

Let us note that this  divergence is gauge-dependent and known. As found in
the early eighties in \cite{Kalashnikov:1980tk,Kajantie:1982xx} and illustrated in the textbook~\cite{Kapusta:2006pm}, the position of the pole at $q\approx 3 g_3^2 N_c T/16$ shifts 
in a generic covariant gauge to $q\approx [8+(\xi+1)^2] g_3^2 N_c T/64\kb{>0}$ \kb{ for all $\xi$}, with $\xi$ the gauge parameter, $\xi=1$ for Feynman gauge. 
Physically, this pole would be
absent for an abelian theory: it occurs for $q\sim g_3^2T$ and arises 
from self-interactions of soft, transverse gluons, hence the proportionality to
$N_c$. As pointed out long ago in~\cite{Kalashnikov:1980tk,Kajantie:1982xx,Heinz:1986kz} and summarized in~\cite{Kapusta:2006pm},
this means that this gauge-dependent resummation has introduced an artificial sensitivity
to this ultrasoft, chromomagnetic scale where  perturbation
theory breaks down~\cite{Linde:1980ts}. 
We recall that, as we mentioned
in the main text, \cite{DEramo:2012uzl} analyzed an altogether
similar problem for transverse momentum broadening and reached
the same conclusions we are drawing here. 
To avoid this spurious pole, they introduced
a scheme that resums the HTL self-energy for $Q\lesssim g_3T$ and the 
full one above. 
\kb{Indeed, if one drops the non-HTL $-3/16 g_3^2 N_c T$ term from the denominator
in Eq.~\eqref{startsalviosanityboomstart}, one recovers the IR-finite, pathology-free
HTL contribution in this region.}

For these reasons, we think that this pathology of the gauge-dependent 
resummation scheme is not unique to axion production --- gravitino 
\cite{Rychkov:2007uq} and axino \cite{Strumia:2010aa} calculations
would also be affected. See also~\cite{Eberl:2020fml,Eberl:2024pxr}
for recent improvements on the gravitino calculations that do not,
however, seem to touch the main issue of the divergence.

We conclude with two final remarks. First, the authors of \cite{Rychkov:2007uq,Salvio:2013iaa} justify their scheme by noting that the HTL spectral function
is a good approximation to the full one-loop one only for $Q\ll T$ 
and should thus not be used for all $Q$. We agree with this only partially.
Let us look at the spectral function of a resummed propagator in the cut regime, such as the one given 
by Eq.~\eqref{spfret}. It features the imaginary part of the self-energy 
at the numerator, whereas the denominator contains both the real and the 
imaginary parts.
\kb{Crucially}, it is only the (imaginary part of the) 
self-energy at the numerator
that must do away with the HTL approximation, so that it can be used for all $Q$.  The denominator can instead be kept
in the HTL approximation, as self-energy resummation is only relevant at soft $Q$,
\kb{where the HTL holds}. Indeed, this corresponds precisely to the 
scheme of \cite{Arnold:2002zm,Arnold:2003zc}, of which our tuned mass scheme is a
simplified implementation. As the previous discussion has shown,
replacing HTL self-energies at the denominator with full, gauge-dependent 
ones leads to well-known pathologies in non-abelian gauge theories.

Finally, this gauge-dependent resummation scheme would not be pathological 
for abelian gauge theories, where the one-loop gauge-boson 
self-energy is actually gauge 
invariant. Provided the problem with the $s$-channel contribution is addressed,
the method of \cite{Rychkov:2007uq,Salvio:2013iaa} would then provide another
scheme that agrees with the strict one at small $g_1$. By the arguments 
of this section and of App.~\ref{app_phase} --- see in particular 
Eqs.~\eqref{ogdiffstart} and Eq.~\eqref{ogdiff} and the text around them --- we 
expect that this scheme would differ from the strict one at relative 
$\mathcal{O}(g_1)$ for a theory with light bosonic (and possibly fermionic)
charge carriers, such as the $U(1)$ component of the SM in the symmetric phase.
For a QED-like theory with fermionic charge carriers only, the difference
would be $\mathcal{O}(g_1^2)$.\footnote{
\kb{Recently, \cite{Danhoni:2024ewq} determined the shear viscosity
of hot and dense QCD at leading- and next-to-leading order. For
baryon chemical potentials sufficiently larger than the temperature,
the one-loop gluon self-energy is dominated by the gauge-invariant quark
loop. The authors thus compared the numerical impact of resumming
HTL and full self-energies in the denominators. In this well-defined and pathology-free case they find that the discrepancy is $\mathcal{O}(10\%)$
or less at intermediate couplings, in full agreement with our argument here.}
} Hence, we do not think that this abelian, amended
version of the method of \cite{Rychkov:2007uq,Salvio:2013iaa} would be 
an any more natural extrapolation to the $g_1\sim 1$ range than our tuned 
scheme, which is, we remark, far lighter from a numerical standpoint.

\section{Theory uncertainty in perturbative thermal QCD}
\label{app_QCD}

The calculation of the thermal axion rate in the KSVZ model is 
essentially a perturbative calculation in hot QCD. In particular, the temperature
range where the reliability of such a calculation is most uncertain
is in the vicinity of the QCD transition. To this end,
we summarize here a few important findings from the hot QCD
and heavy-ion collision literature that are directly relevant to
our case. We refer to \cite{Ghiglieri:2020dpq} for a comprehensive review
of perturbative hot QCD.

As mentioned, the axion rate is strongly connected with the transverse
momentum broadening coefficient $\hat{q}$. Indeed, they are both 
part of a broader class of \emph{dynamical observables}, to be 
opposed to \emph{thermodynamical ones}. The latter, such as the pressure,
energy density, entropy, etc., are given by thermal expectation values
of local operators, e.g. the energy-momentum tensor. The 
former, on the other hand, are given by operators that are non-local
in time, such as the retarded two-point function giving the axion
rate in Eqs.~\eqref{defgammapi} and \eqref{defpi}.

This important distinction makes it so that thermodynamical observables can readily be evaluated in compactified Euclidean spacetime. They are 
amenable (at vanishing baryon chemical potential) to precise lattice 
determinations and to high-order perturbative ones. As summarised e.g.
in \cite{Ghiglieri:2020dpq}, perturbative determinations of the pressure 
through sophisticated resummations and Effective Field Theories reach $\text{N}^4\text{LO}$ 
and are in reasonable agreement, though with large uncertainties, with lattice QCD at 
temperatures above twice the pseudo-critical one. 
These perturbative determinations also show the role that physics at or below
the soft $g_3T$ scale plays in the apparent slow convergence of the expansion.
It bears repeating that no
gauge-dependent resummations of the full one-loop gluon self-energies are performed in these calculations, 
as the community is aware of the pathologies that arise, as discussed in the previous appendix.

For what concerns dynamical quantities, much less is known. Direct
lattice determinations are not possible, as these quantities are intrinsically Minkowskian and require a highly non-trivial analytical 
continuation from Euclidean space-time, in the form of an ill-posed inverse problem --- see e.g. \cite{Meyer:2011gj} for a review. 
Perturbative determinations too are more challenging, with most quantities
being known to leading order. NLO corrections, of relative
$\mathcal{O}(g_3)$, have been determined only for a handful of 
observables, as reviewed by one of us in \cite{Ghiglieri:2020dpq}.
These include the momentum broadening coefficient of heavy quarks in \cite{CaronHuot:2007gq,Caron-Huot:2008dyw}, 
that of light quarks and gluon --- $\hat{q}$ --- in \cite{CaronHuot:2008ni},
the photon production rate in \cite{Ghiglieri:2013gia}, the 
shear viscosity in \cite{Ghiglieri:2018dib} and the isotropic thermalisation time in \cite{Fu:2021jhl}.

As summarised briefly in \cite{Ghiglieri:2022sui}, observables that 
are directly sensitive to the physics of transverse momentum broadening,
such as $\hat{q}$ itself and the shear viscosity, show 
large $\mathcal{O}(g_3)$ corrections, which overtake the LO
at $g_3\sim 0.5$.  On the other hand, processes that are not directly
sensitive to that physics, such as  photon production
and isotropic thermalisation, have $\mathcal{O}(g_3)$ corrections at the $30\%$  level
even when extrapolating to $\alpha_s\sim 0.3$ ($g_3\sim 2$) which is
the expected value in the vicinity of the QCD crossover.
As we mentioned, one possible explanation could lie in an 
``overscreening'' in the leading-order formulation of transverse
momentum broadening  \cite{Muller:2021wri}. 

We further remark that the NLO calculation of the shear viscosity in
\cite{Ghiglieri:2018dib} included many different $\mathcal{O}(g_3)$ 
contributions, coming from
several different momentum regions and physical mechanisms. One of the 
main findings was that  NLO corrections to $\hat{q}$ dwarf all 
others. This motivates our NLO
Ansatz as a proxy for  conservatively ``very large'' NLO contributions
and further suggests how important transverse-momentum-broadening physics
might not be captured in the ``overscreened'' LO formulation of $\hat{q}$.
We also mention that non-perturbative, all-orders determinations of 
the soft, classical contribution to transverse momentum broadening
in \cite{Panero:2013pla,Moore:2019lgw,Moore:2021jwe,Schlichting:2021idr}
through dimensionally-reduced lattice field theory point to a smaller
value for $\hat{q}$ than that obtained at NLO. 

Finally, we remark that our methodology of comparing several 
LO-equivalent determinations of a dynamical quantity is well rooted
in the hot QCD literature. For instance, when computing the 
shear viscosity and quark number diffusion to leading order, the authors of \cite{Arnold:2003zc}
compared results obtained by resumming HTL self-energies in the 
naively divergent denominators with other LO-equivalent schemes where
HTL resummation is switched off above an exchanged three- or four-momentum
of the order of the temperature --- see Fig.~4 there, finding 
an $\mathcal{O}(1)$ uncertainty for $\mD/T\approx 2$.
\cite{CaronHuot:2007gq,Caron-Huot:2008dyw} compared  strict and tuned-like
LO schemes with their new NLO determination of heavy-quark momentum broadening.
More recently,
 \cite{Boguslavski:2024kbd} compared results for non-equilibrium
dynamics from the QCD kinetic theory obtained from their implementation
of a tuned scheme and from the HTL-resummed denominator, as per \cite{Arnold:2002zm}. In particular, their Sec.~V.G 
and Fig.~10 explicitly consider the case of $\hat{q}$, finding a small
discrepancy between the two.


\bibliographystyle{JHEP}
\bibliography{ref.bib}

\end{document}